\documentclass[twocolumn, superscriptaddress]{revtex4-2}

\usepackage{cancel}
\usepackage{graphicx}
\usepackage[braket, qm]{qcircuit}
\usepackage{braket}
\usepackage{changes}

\usepackage{dcolumn}
\usepackage{bm}
\usepackage{xcolor}
\usepackage{svg}
\usepackage{amsmath}
\usepackage{hyperref}

\usepackage{afterpage}
\usepackage{adjustbox}

\usepackage{rotating}

\begin{document}

\title{Low Depth Virtual Distillation of Quantum Circuits by Deterministic Circuit Decomposition}

\author{Akib Karim}
\email{akib.karim@csiro.au}\affiliation{Data 61, CSIRO, Research Way Clayton 3168, Victoria, Australia}
\author{Shaobo Zhang}\affiliation{School of Physics, University of Melbourne, Parkville 3010, Victoria, Australia}
\author{Muhammad Usman}\affiliation{Data 61, CSIRO, Research Way Clayton 3168, Victoria, Australia}\affiliation{School of Physics, University of Melbourne, Parkville 3010, Victoria, Australia}

\begin{abstract}
Virtual distillation (VD) using measurements of multiple copies of a quantum circuit have recently been proposed as a method of noise mitigation of expectation values. Circuit decompositions known as B gates were found only for 1-local Hamiltonians however practical problems for chemistry require n-local Hamiltonians which cannot be corrected with B gates. We discover low depth circuit decompositions for expectation values for n-local Pauli strings by combining multiple projections to recover the correct measurement statistics or expectation values. Our method adds linear entangling gates with number of qubits, but requires extra measurements. Furthermore, in applications to find ground states such as the variational quantum eigensolver (VQE) algorithm, the variational principle is required which states the energy cannot go below the ground state energy. We find that the variational principle is violated when using B gates and is preserved if using our low depth decomposition on all expectation values. We perform demonstration on real devices and demonstrate our decomposition can mitigate real experimental noise in VQE for the H$_2$ molecule with a two qubit tapered mapping, H$_3$ with three qubits, and H$_2$ with four qubits. Our decomposition provides a way to perform duplicate circuit virtual distillation on real devices at significantly lower depth and for arbitrary observables.
\end{abstract}

\maketitle

\section{Introduction}

Quantum computing is a rapidly emerging field that is anticipated to run algorithms capable of solving certain computationally hard problems with improved scaling to its classical counterpart~\cite{Montanaro2016,Shor1994}. Unfortunately, current quantum devices are in the era of Noisy Intermediate Scale Quantum (NISQ) because they suffer from significant noise, which reduces accuracy of general algorithms or renders them useless for quantum advantage. However, NISQ specific algorithms have been designed where quantum advantage has thought to have been found, for example, boson sampling to calculate matrix permanents~\cite{White2013}, increased information capacity for finite use of an amplitude damping channel~\cite{Akib2018}, and the ability to generate statistics from random quantum circuits~\cite{Bremner2017}, however whether specific implementations show advantage is a matter of debate~\cite{Arute2019,Pan2022,bremner2023}. For algorithms with more widespread applications, advantage is only reached with significantly more qubits and less noise. For example QAOA, which can calculate various graph theory problems like MAXCUT, is thought to need 420 qubits for supremacy given some complexity assumptions~\cite{Dalzell2020}. Quantum error correction can guarantee that if noise is below a threshold, multiple physical qubits can be mapped to correctable logical qubits and the error rate can be bound arbitrarily~\cite{Shor1996}. These Fault Tolerant devices will be able to show quantum advantage without modifying the algorithms to consider noise. However, with our current NISQ devices, we require more specialised algorithms.

Some NISQ algorithms are considered tolerant to specific errors, such as the Variational Quantum Eigensolver (VQE) for finding ground state energies of Hamiltonians~\cite{Peruzzo2014}. VQE uses a parametrised ansatz circuit and measures the energy of the resultant state. The parameters are adjusted until the minimum is found and this compensates for errors due to parameter setting or gate calibration. However, VQE cannot compensate for incoherent errors that result in mixed states and there is a need to design noise mitigation techniques for these incoherent errors.

\begin{figure*}
    \centering
    \includegraphics[width=\textwidth]{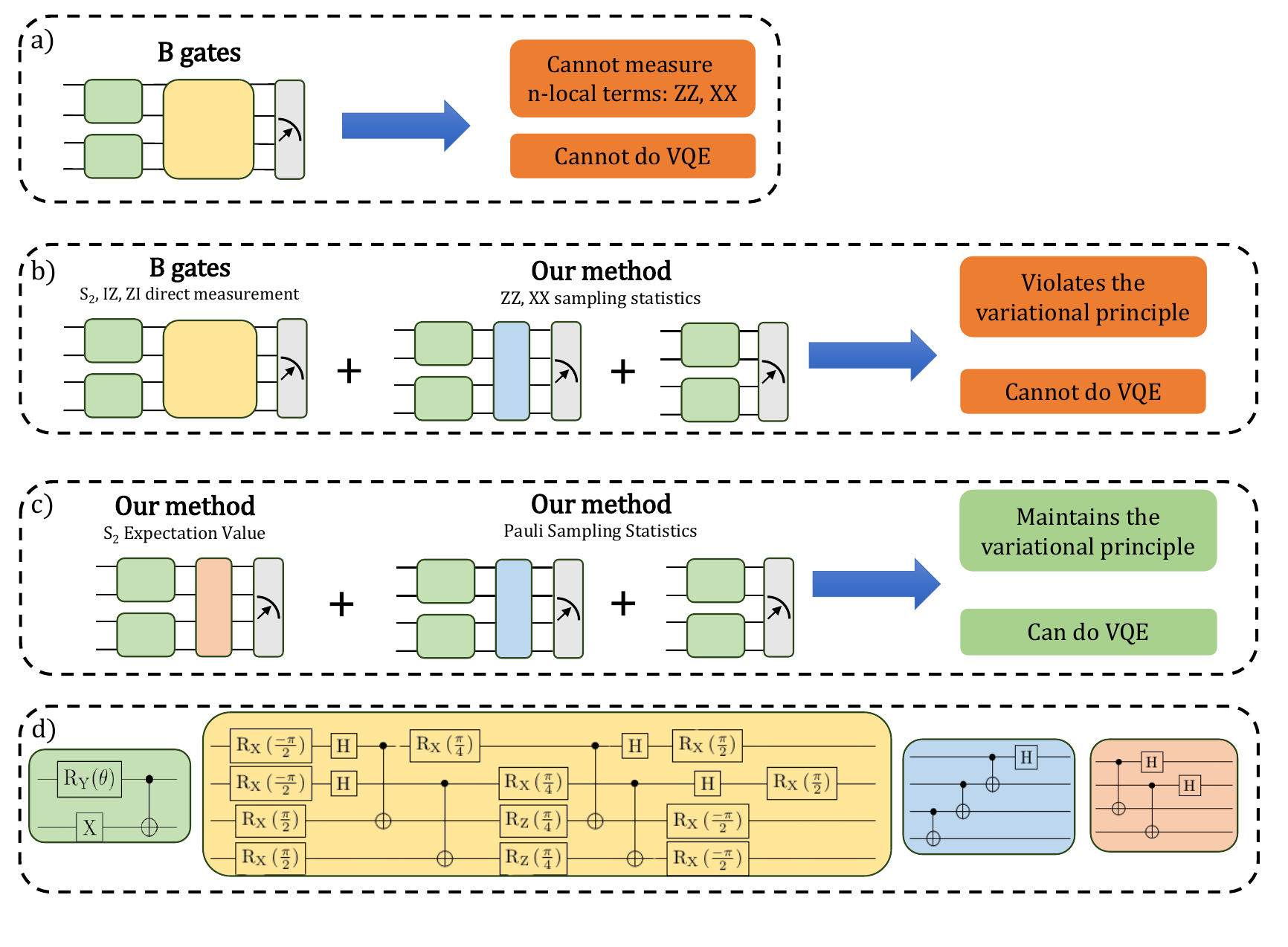}
    \caption{Diagram of virtual distillation circuit implementations for a two qubit H$_2$ VQE calculation. a) is the original B gate proposal from Huggins et al.~\cite{Huggins2021}, which is only able to measure 1-local Hamiltonian terms and unable to perform the VQE calculation. In b) for 1-local terms, we use B gates and for 2-local terms, we use our novel method of decomposition where the measurement statistics can be recreated with an entangling gate and computational measurement instead of direct measurement. This results in the variational principle being violated and VQE is unable to be performed. c) We replace the B gates with our method to recreate the expectation value of $S_2$ using an entangling projection and computational measurements. The Pauli terms are recreated with the same computational measurement and an entangling projection. d) shows the circuits for Ansatz, B gates, the entangling projection to recreate Pauli sampling statistics, the entangling projection to recreate the $S_2$ expectation value}
    \label{fig:enter-label}
\end{figure*}

Many existing strategies of incoherent noise mitigation rely on characterisation of the device such as Bayesian Readout Error Mitigation~\cite{Nachman2020}, Probabilistic Error Correction~\cite{Temme2017}, Random Circuit Sampling~\cite{Endo2018}or machine-learning based methods~\cite{Sack2023}. Similarly, if the noise model is known analytically, there exist solutions to measure and undo their effect~\cite{Endo2018,Urbanek2021}. While these have shown success, these characterisation methods become quickly outdated due to noise drift in real devices and methods without this assumption are needed in addition. Zero Noise Extrapolation has found success lately by using additional measurements on circuits of increasing noise to extrapolate back to a theoretical noise-free value~\cite{Temme2017}. However, this can lead to a biased estimator depending on the noise amplification method~\cite{Kim2023}. There also exist error detection and postselection methods which increase number of measurements to select samples with less noise~\cite{Mezher2022}. 

Recently, a method of virtual distillation was designed that uses a duplicate copy of the circuit to correct expectation values after measurement~\cite{Huggins2021}. 

This found success with small systems that used only 1-local expectation values, but they provided no general scheme for larger systems that also require n-local expectation values. The extension of the B gate scheme would require a unique circuit for each expectation value with no bound on circuit depth. As the depth of the noise mitigation circuit increases, it can add more noise than is removed, rendering it useless.

Subsequently, further papers that used virtual distillation only measured observables that can be rewritten as 1-local expectation values~\cite{google2022}, by calculating a density matrix from maximum likelihood~\cite{Cao2023}, or shadow tomography methods rather than direct measurement, which will require exponential measurements for arbitrary states~\cite{Alireza2023}. Recently, a method of circuit cutting for the swap gates of the B gate circuit was found so that the correcting gates can be simulated on a classical computer, but this also requires probabilistic reconstruction~\cite{li2023}. There is a need to devise new methods of deterministic circuit decomposition to enable duplicate circuit virtual distillation for use in practical calculations.

In this work, we propose a method of recovering the statistics of the proposed unitary projections in Ref.~\cite{Huggins2021} by combining measurements from computational and specific entangling projections. We find methods to reconstruct the output sampling statistics or reconstruct the expectation value. Our method ensures linear scaling maximum additional depth. In particular, we find only $n$ additional two-qubit entangling gates are required for a circuit with $n$ qubits before duplication. We devise a circuit decomposition that reconstructs the sampling statistics as well as a decomposition that reconstructs the expectation value. We show that a combination of these allows the variational principle to be retained while also having lower depth than the B gates. In the appendix, we show alternative methods useful for chip topologies such as: measuring Pauli strings by reconstructing the expectation value, which increases the depth for fewer measurements; or direct decomposition of observables to Pauli strings which has no extra entangling gates but exponentially more measurements for systems where shots are less costly than depth.

Our method works for any algorithm that requires expectation values, therefore we demonstrate its effectiveness on the variational quantum eigensolver (VQE)~\cite{Peruzzo2014} for two qubits (H$_2$), three qubits (H$_3$), and four qubits (H$_2$ no symmetry reduction). We show noise resilience using simulations as well as by running demonstrations on IBM quantum devices through cloud-based access.

\section{Theoretical Framework}
For the trial state $\rho$ and the Pauli string $O$, the expectation value is given by:
\begin{equation}
    tr(O\rho).
\end{equation}

Normally, we rotate $\rho$ such that $O$ is diagonalised, measure in the computational basis and combine the counts based on the corresponding eigenvalues, i.e. if $O = UDU^\dagger$, where $D$ is a diagonal matrix and $U$ is a unitary matrix then the measurement becomes:

\begin{equation}
    tr(UDU^\dagger \rho) = tr(DU^\dagger \rho U), \label{eq:U}
\end{equation}
since the trace is invariant under cyclic permutations. 

If there are errors, we can virtually purify the density matrix by replacing these measurements with:

\begin{equation}
    \frac{tr(O\rho^{2})}{tr(\rho^{2})}.
\end{equation}

It has been shown in Ref.~\cite{Huggins2021} that this quantity can be calculated using two copies of the circuit to create the density matrix $\rho^{\otimes2}$ and measuring 

\begin{equation}
    \frac{tr(OS_{2}\rho^{\otimes2})}{tr(S_2\rho^{\otimes2})},\label{eq:fracti}
\end{equation}
where $S_{2}$ is the operator that swaps each qubit with their duplicate. It should be noted that $Tr(S_{2}\rho^{\otimes2})$ is the definition of the purity of $\rho$.

For general observables we construct the symmetrised operator: 

\begin{equation}
O^{sym} = \frac{O + S_{2}O}{2}. \label{eq:O}
\end{equation}

This is the expectation value of $O$ on the original circuit added to the expectation value on the duplicate circuit. This is desirable as the null space of the new operator is much larger as it consists of the states where both the circuit and duplicate circuit have zero eigenvalue, but also when they differ by a sign. The components of a state in this null space will not contribute to the expectation value and can be ignored when reconstructing the measurements.

Huggins et al.~\cite{Huggins2021} found that 1-local observables, i.e. $O = ZII...$, commutes with $S_{2}$ and therefore, they can construct one unitary ($U$ from equation~\ref{eq:U}) to project on to measure the entire quantity in equation~\ref{eq:fracti}. Their specific circuit decomposition is called the B gate.

Unfortunately, multi-qubit operators do not commute with $S_{2}$ in general. We still need the B gate unitary to measure $tr(S_{2}\rho^{\otimes2})$, but now we need a new matrix that rotates the state to measure the numerator. We can naively decompose this matrix into circuits using general decomposition algorithms, however this is incredibly deep and requires a new circuit for every set of commuting Pauli strings.

To enable use for arbitrary systems, we require a general decomposition scheme for all Pauli strings. We first investigate the eigenvectors of the observables. For $S_{2}$, the eigenvectors will be the states invariant to swapping. For example, consider a four qubit state $\ket{abcd}$, $a,b,c,d \in \{0,1\}$. The effect of $S_{2}$ is:
\begin{equation}\label{eq:eigenvects}
    S_2\ket{abcd} = \ket{cdab}.
\end{equation}

The eigenvectors are therefore:
\begin{enumerate}
    \item $\ket{abcd}$, for $a=c;b=d$ or
    \item $\frac{1}{\sqrt{2}}(\ket{abcd} + \ket{cdab})$ otherwise.
\end{enumerate}

The computational basis states are exclusively either already eigenvectors or can be rotated into the eigenbasis by superposition with their swapped basis. The technique from Huggins et al.~\cite{Huggins2021} finds a single unitary that can project on the superposition states while leaving the rest in the computational basis. We show that we can recreate the same statistics by combining measurements on computational basis and entangled bases with less depth.

We now show how to decompose this superposition basis into CNOT and Hadamard gates. Depending on the native entangling gates of the hardware, further optimisations can be made. We choose these as they are a subset of Clifford gates, which can be simulated efficiently classically and allow for noise mitigation schemes that characterise noise by comparing Clifford gate simulations to experiment. 

\subsection{Reconstructing Sampling Statistics for Pauli Strings}
For a general four qubit state $\ket{abcd}, a,b,c,d \in \{0,1\}$, the effect of applying a CNOT with control qubit $0$ and target qubit $1$ then a Hadamard on qubit $0$ is:
\begin{equation}\label{eq:HCNOT}
    H_0~CNOT_{0,1} \ket{abcd} = \ket{0bcd} + XXII\ket{0bcd},
\end{equation}

where the subscript labels the qubits. We therefore create the superposition of the first qubit in $\ket{0}$ and $\ket{1}$ and the second qubit identity or bit flipped. Using this, we can decompose the superpositions in equation~\ref{eq:eigenvects} into equivalence classes of the form in equation~\ref{eq:HCNOT}. For a duplicate circuit of $2n$ qubits, the number of 
total equivalence classes is $2^n$, but they only all need to be measured for $\langle S_2\rangle$, which is one reason we find an alternative method for specifically $S_2$ below. For observables, only a subset need to be measured because we use the symmetrised form in equation~\ref{eq:O}. If the expectation values of $O$ and $S_2 O$ differ by a sign, the value becomes $0$. Furthermore, since Hamiltonians can be decomposed into multiple Pauli strings, they may share equivalence classes. Any technique that can remove the need to measure strings of one class will reduce measurements. For example, Pauli grouping to find a unitary to maximise simultaneous measurement of Pauli strings could be used to reduce the number of string classes and is compatible with our decomposition as we recreate sampling statistics.

The final procedure is to then: identify the set of entangling projections necessary for the specific Paulis required; measure those and the computational basis; directly reconstruct the measurement vectors where statistics from states invariant to swap are taken from computational measurements and the rest are taken from their respective entangled measurement; normalise the measurement vector; calculate relevant Pauli expectation values. 

We note that our circuit for the 1-local measurements has significantly lower depth than the B gate proposed in Ref.~\cite{Huggins2021}. 

\subsection{Recreating the Expectation Value of $S_2$}
Rather than recreate the sampling statistics, we can instead decompose the expectation value directly. Previously, we could normalise the sampling statistics, but this cannot be performed with only the expectation value. Physicality requires the expectation value to fall in the range $[-1,1]$ and for $\langle S_2 \rangle$ this is not a problem as it is only close to $1$ with no noise.

Consider $\ket{\psi} = a_0 \ket{00} + a_1 \ket{01} + a_2 \ket{10} + a_3 \ket{11}$. We can expand the expectation value $\bra{\psi}S_2\ket{\psi}$ as: 
\begin{equation}
    a_0 a_0^\dagger \bra{00} S_2 \ket{00} + a_0 a_1^\dagger \bra{00} S_2 \ket{01} +... etc.
\end{equation}

We can group these into our two sets of basis states from equation 6. If the basis state $\ket{i}$ is invariant, then they will contribute terms of the form $|a_i|^2$, whereas if a pair of basis sets, $\ket{i}$ and $\ket{j}$, are mapped to each other by $S$, then they contribute $a_i a_j^\dagger + a_j a_i^\dagger$. 

In order to calculate the full expectation value, we consider a circuit where we apply CNOTs on pairs of qubit copies (qubits labelled by subscript) and Hadamards on all qubits of one copy. For a two qubit circuit this is:

\begin{equation}
(H\otimes H\otimes I \otimes I) CNOT_{0,2} CNOT_{1,3}.
\end{equation}

When applied to a computational basis state $\ket{abcd}$ this creates a superposition over four states 
\begin{equation}\label{eq:states}
    IIII\ket{00cd} \pm IIIX\ket{01cd} \pm IIXI\ket{10cd} \pm IIXX\ket{11cd},
\end{equation} where the sign depends on the initial state. Importantly, for any number of qubits, it will contain the original computational state and the generator Pauli strings calculated above, however, it will also contain the superposition of additional states. Consider an arbitrary state $\sum_{abcd} A_{abcd} \ket{abcd}$. Applying the circuit then measuring in the computational basis will give outcomes that are linear combinations of all four states in equation~\ref{eq:states}. It is always possible to find a linear combination of outcomes that recreate the expectation value for $S_2$ because it is simply the addition of terms where the states are swapped. Appendix A details how to find the linear combination of outcomes with examples for one and two qubits. Appendix B then describes how to use this method for measuring observables, which requires multi-control Z gates.

The full procedure is: first calculate the linear combination of outcomes that eliminates the additional cross terms; then measure in computational and this one extra pairwise entangling projection; the expectation value of $S_2$ is the addition of the computational measurements for states invariant to $S_2$ with the linear combination of outcomes from the entangled projection. We note that the pairwise entanglement is identical to the B gate and requires only one entangling gate per pair compared to B gate which needs two. It also requires no rotations and uses only Clifford gates.

\section{Demonstrations on IBM devices}

To benchmark the performance of our methodologies, we run our circuits on IBM superconducting devices. It should be noted in this paper, we report all states in conventional (big endian) format, which is reversed from IBM measurement output.

For two qubits, we generate a Hamiltonian for the H-H molecule at a fixed distance using PySCF~\cite{Sun2017} with a STO-3G basis~\cite{Pople2003}. This is mapped to a two-qubit basis using the parity mapping with Z$_2$ symmetry reduction. We then use a manually simplified UCCSD Ansatz with one parameter, $\theta$, given by preparation in the computational basis, $Ry(\theta)\otimes X~CNOT$ and measurement in the computational basis. As there is only one parameter, we sweep from $[-\pi,\pi]$ with $51$ steps. This is repeated with auxiliary error correcting circuits. For each Hamiltonian, the angle that minimises the energy for the raw data was found and the corrected values were calculated for that angle. Demonstrations were performed on $ibm\_hanoi$ using a mapping to qubits 1,4,7,6 as indicated in Appendix F.

\begin{figure}
    \centering
    \includegraphics[width=\columnwidth]{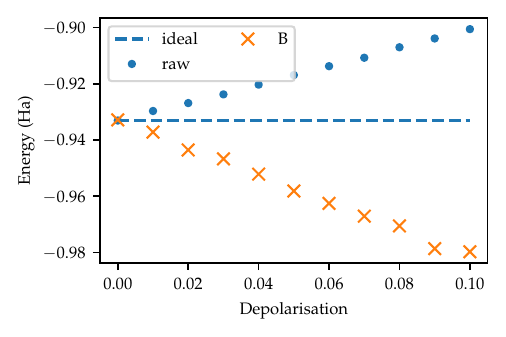}
    \caption{H$_2$ energy at distance of $2$\AA~as depolarisation noise on CNOT gates increases. Ideal refers to the noise free value at $0$ depolarisation. Raw refers to directly measuring the ansatz, B refers to using B gates~\cite{Huggins2021} for VD and violates the variational principle by being below ideal.}
    \label{fig:H2-var}
\end{figure}

\begin{figure}
\flushleft{(a)}
\includegraphics[clip,width=\columnwidth]{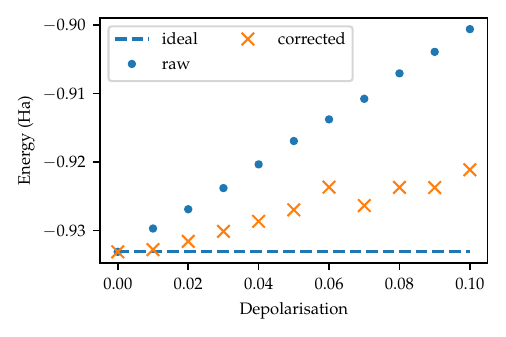}

\vspace{-1.1cm}
\flushleft{(b)}
\includegraphics[clip,width=\columnwidth]{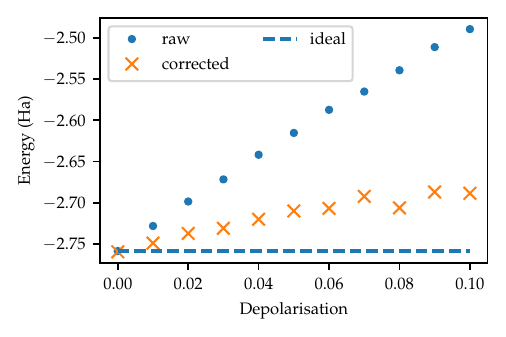}

\vspace{-1.1cm}
\flushleft{(c)}
\includegraphics[clip,width=\columnwidth]{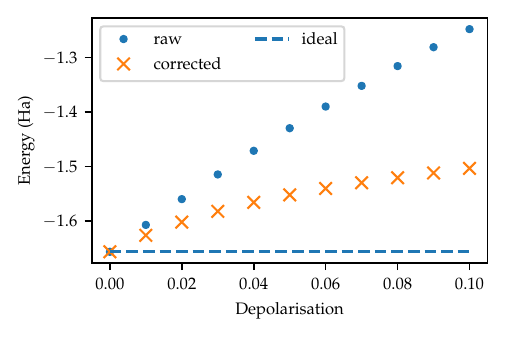}
\caption{Effect of depolarisation on ground state energy for 2 qubit (H$_2$), 3 qubit (H$_3$), and 4 qubit (H$_2$ no symmetry reduction) systems at a distance of $2$\AA. Simulated data from depolarisation channel on each CNOT gate. Dotted blue line gives ideal noiseless value. Blue dots show raw measured energy. Yellow crosses give corrected energy.}
\label{fig:fake}
\end{figure}

Similarly, for three qubits, we used a linear chain H-H-H with equal distances between Hydrogens. We used STO-3G with the Jordan-Wigner mapping and qubit tapering~\cite{bravyi2017} to get a three-qubit basis. A hardware efficient ansatz with nine parameters was used given by two layers of Ry, CNOT, Ry, CNOT, Ry. Calculations confirming the ability of the ansatz to find the ground state can be found in Appendix D. Optimal values were found with simulated data using a provided noise model as performed in Li et al.~\cite{li2023}. The optimal circuit was then measured in computational basis and on entangled corrective projections. Demonstrations were performed on $ibm\_hanoi$ using a mapping to qubits 5,3,2,1,4,7 as indicated in Appendix F. 

Finally, for four qubits, we used the H-H molecule with STO-3G and Parity mapping but without any symmetry reduction. We used a hardware efficient circuit with Ry, Rz, CNOT, Ry, Rz, CNOT, Ry, Rz layers across all qubits.This ansatz is able to find the ground state for the Hamiltonian with Parity mapping but not Jordan-Wigner mapping as detailed in Appendix D. Optimal values were found using a provided noise model using simulated data. The optimal circuit was then measured in computational basis and on entangled corrective projections. Experiments were performed on $ibm\_hanoi$ with mapping to qubits 1,2,3,6,4,7,5,8 as indicated in Appendix F.

The entangled projections are determined to recreate sampling statistics for each Pauli string and pairwise entangling gates to recreate the expectation value of $S_2$.

\section{Results and Discussion}

\begin{figure}[t!]
\flushleft{(a)}
\includegraphics[width=\columnwidth]{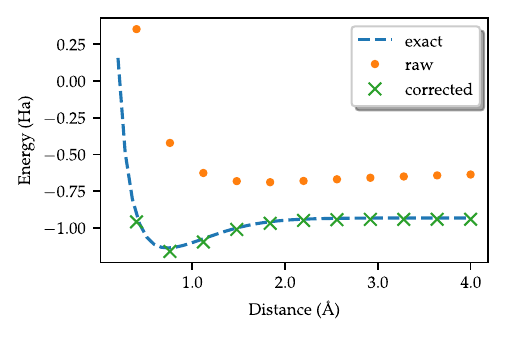}

\vspace{-1.1cm}
\flushleft{(b)}
\includegraphics[width=\columnwidth]{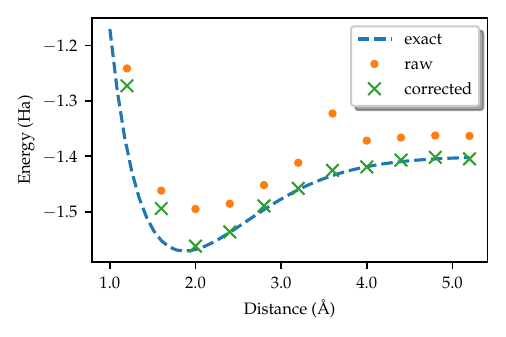}

\vspace{-1.1cm}
\flushleft{(c)}
\includegraphics[width=\columnwidth]{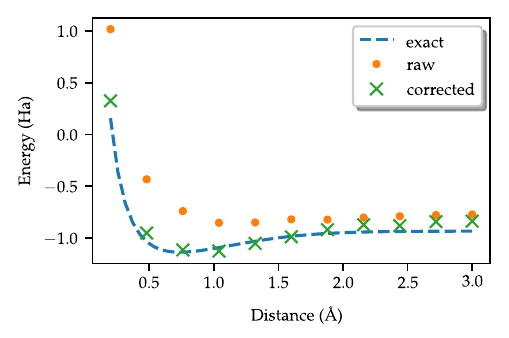}
\caption{Dissociation curves for two-qubit, three-qubit, and four-qubit systems on $ibmq\_hanoi$. Exact refers to direct diagonalisation of the Hamiltonian after qubit mapping. Raw refers to directly measuring the ansatz. Corrected refers to using VD with our circuit decomposition. a) shows H$_2$ molecule with UCC ansatz and parity reduction. b) shows H$_3$ linear chain molecule with hardware efficient ansatz. c) shows H$_2$ molecule with hardware efficient ansatz.}
\label{fig:real}
\end{figure}

We first perform a calculation with B gates from Ref.~\cite{Huggins2021} as a comparison to previous work. We will use the two-qubit H$_2$ molecule calculation as an example, which has a Hamiltonian that decomposes to $IZ$,$ZI$,$ZZ$,$XX$ strings. Since the B gate cannot calculate multi-qubit measurements, we must use the method developed in this work for reconstructing the sampling statistics for $ZZ$ and $XX$ Paulis.

Figure \ref{fig:H2-var} gives the measured ground state energy of H$_2$ at $2$Å distance as depolarisation on each two qubit gate increases. The depolarisation parameter ranges between $0$ at no noise and $1$ where the qubit is in the maximally mixed state. The dotted line refers to the ideal noiseless measurement; the blue dots give the raw measurement without any correction; the yellow crosses give the B gate corrected values.

It is obvious that depolarisation noise causes the B gates to immediately violate the variational principle. This is because it has four CNOTs to calculate $S_2$, whereas our work only uses two at most. This means $S_2$ sees four depolarisation channels and underestimates the purity. 

O'Brien et al.~\cite{google2022} also noted violation of the variational principle with virtual distillation, however they bound expectation values between [-1,1] which they state is clearly qualitatively incorrect. They are correct in that assessment: it is obvious that if $S_2$ measurements have more error than Pauli measurements, observables will appear to take their maximum value when the noiseless value may be nowhere near it. We will now provide results for the method developed in our work which takes this into account and ensures we can retain the variational principle.

Figure~\ref{fig:fake} shows the same two-qubit CNOT gate depolarisation noise as in Figure~\ref{fig:H2-var}. Figure~\ref{fig:fake} (a) shows the same two-qubit H$_2$ system and the raw measurement given by blue dots is the same as in Figure~\ref{fig:H2-var}. Figure~\ref{fig:fake} (b) shows the three-qubit H$_3$ linear chain molecule and Figure~\ref{fig:fake} (c) shows the four-qubit H$_2$ molecule with no symmetry reduction.

Firstly, our method does not violate the variational principle, as the computed energy with noise is always higher than the noiseless ideal dotted blue line. Secondly, the corrected values are always a lower energy than the raw values, which means our method is successfully mitigating noise. Specifically, it means the additional entangling gates do not add more noise than they cancel out. We note that while the uncorrected raw energies increase linearly, in this noise range, the corrected values increase much slower, especially for four qubits in Figure~\ref{fig:fake} (c). 

Finally, it is observed that the larger systems are more sensitive to small noise. Around $0.01$ depolarisation, the two qubit system is able to be corrected to almost ideal, whereas the four qubit system cannot. In comparison, due to the slow increase of the corrected values to noise, at high noise values, larger systems perform well. VD by itself is therefore optimal at relatively higher noise (up to reasonable noise for a current device). However, since it always corrects, there are other techniques that can be used in conjunction that could mitigate VD at small noise and allow it to reach the ideal case.

After a basic test of our method in noisy simulation environments, we can now implement it on a noisy IBM quantum device. We calculate the full dissociation curve for the molecules in question as distance increases. The results are given in Figure~\ref{fig:real}. The ideal curve is given as a dotted blue line, the raw values are yellow dots, and the corrected values are green crosses. As above, Figure~\ref{fig:real} (a) is the two qubit H$_2$ calculation, (b) is the three qubit H$_3$ linear chain, and (c) is the four qubit H$_2$ calculation with no symmetry reduction. 

The results follow similar trends to the simulated data. For the smallest system, in Figure~\ref{fig:real} (a), we note that the ideal energy can be recreated up to shot noise, despite large errors on the raw data. For the largest system in Figure~\ref{fig:real} (c), we can note similar correction close to ideal, however the curve as the hydrogen atoms are separated is not able to recreate the exact energies. Furthermore, from the comparison of raw and exact data, it appears that the noise in Figure~\ref{fig:real} (c) is much less than in Figure~\ref{fig:real} (a). 

Similarly, the effect of shot noise was considered. The demonstrations from Figure~\ref{fig:fake} were simulated with $8196$ shots and no other noise. This was repeated $100$ times and the standard deviations in the energy recorded. The standard deviations for the corrected energies are $1.296 \times 10^{-3}~$Ha for two qubits; $1.386\times10^{-3}~$Ha for three qubits; and $2.038\times10^{-3}~$Ha for four qubits. These are significantly smaller than the discrepancy between all measured and exact energies, which shows incoherent noise must be responsible rather than shot noise. However, shot noise is of order of chemical accuracy of approximately $1.594\times10^{-3}~$Ha and this means significantly more shots are required in applications to reach chemical accuracy even if incoherent noise can be mitigated. 

While this method is shown to work, care must be taken to know the exact hardware implementation to ensure that the $S_2$ measurements do not have significantly more error than the Pauli measurements. In particular, additional SWAP gates or dynamical decoupling steps may be introduced by a compiler that can violate this. In the appendix, we detail several alternative methods that can be used to tailor this method to hardware requirements. We show how to calculate Paulis by recreating their expectation value (as we did with $S_2$ in the main text), however this adds a multi-controlled-Z gate, which can be prohibitive and if three-qubit gates can be implemented, the ancilla based Fredkin approach may be more desirable. Similarly, we show how to decompose all expectation values into Pauli strings, which requires no extra entangling gates and has the least noise, but will require the most measurements. At worst, it will need to measure all Paulis for all doubled qubits, at which point it will take fewer measurements to do a tomography on the original circuit and analytically calculate powers of the density matrix. In that sense, for specific problems and hardware limitations, these alternate methods could be useful, however in the specific case of chemical calculations on IBM architecture our method of recreating Pauli sampling statistics and $S_2$ expectation values was found to be optimal.

In conclusion, we have demonstrated that the B gate circuit decomposition for VD only works for observables that commute with $S_2$ and can violate the variational principle otherwise. We found a method for circuit decomposition that maintains the variational principle which enables VQE applications. Our method increases number of measurements but reduces the number of entangling gates compared to B gate method. We found that it performs well at different noise scales and on real devices. Future work can investigate combining VD with other noise mitigation strategies to see if chemical accuracy can be reached. 

\section{Acknowledgements}

The research was supported by the University of Melbourne through the establishment of the IBM Quantum Network Hub at the University.

\bibliography{apssamp}
\section{Appendices}

\subsection{Linear combination of measurements for $\braket{S_2}$}

Recall that the goal is to recreate the expectation value of the swap operator. Let's say we have $\ket{\psi} = A\ket{00} + B\ket{01} + C\ket{10}$, then

\begin{equation}\label{eq:Sappend}
    \bra{\psi}S_2\ket{\psi} = |A|^2 + \overline{B}C + \overline{C}B
\end{equation}

The squared terms are probabilities measured from computational measurement. The goal is thus to measure a quantity that corresponds to the cross terms added together with their complex conjugates. This will be accomplished by the CNOT H circuit (written in circuit order). We want to know what probabilities occur after an arbitrary state $\ket{\psi}$ is projected by the circuit CNOT H on each pair. Ultimately we will be taking probabilities from measuring in the computational basis and taking linear combinations of them. Therefore, the first step is to see what the projection of CNOT H looks like on an arbitrary state $\ket{\psi}$,

In order to make analysis easier, we reorder qubits such that each qubit is adjacent with its corresponding qubit from the copied system. i.e. for two qubit system $\ket{abcd} \rightarrow \ket{acbd}$. This now makes the swap operator $S_2$ act as:

\begin{equation}
    S_2 \ket{abcd...} = \ket{badc...}.
\end{equation}

In this qubit ordering, the projection circuit is now CNOT H on each pair in order, so we can consider the action only on separable pairs of qubits, specifically now only the top two qubits. The operation of CNOT H has form:

\begin{equation}
    \ket{+}\ket{0} \otimes I + \ket{-}\ket{1} \otimes X.
\end{equation}

We want to know what the outcome of measuring in the computational basis after this looks like. If we project onto state $\bra{01}$, we get 
\begin{equation}
    \bra{01} H CNOT = \frac{1}{\sqrt{2}} \bra{01} + \bra{10},
\end{equation}

which means that if we start with the state $\ket{\psi}$ and project using the circuit, we will measure the outcome $01$ with probability:

\begin{equation}
\begin{split}
    \frac{1}{2}(\braket{\psi|01} + \braket{\psi|10})(\braket{01|\psi} + \braket{10|\psi}) \\= \frac{1}{2}(|B|^2 + |C|^2 + \overline{B}C + \overline{C}B),
    \end{split}
\end{equation}

similarly, the outcome $10$ occurs with probability:

\begin{equation}
\begin{split}
 \frac{1}{2}(\braket{\psi|01} - \braket{\psi|10})(\braket{01|\psi} - \braket{10|\psi}) \\= \frac{1}{2}(|B|^2 + |C|^2 - \overline{B}C - \overline{C}B),
 \end{split}
\end{equation}

Therefore, if we subtract the probability of measuring outcome $10$ from $01$, we get $\overline{B}C + \overline{C}B$, which matches the quantity required in equation~\ref{eq:Sappend}.\\

Explicitly, for arbitrary numbers of qubits, the action of CNOT H on the top two qubits is to map:
\begin{equation}
\begin{split}
    \bra{00}\otimes\bra{abc...} \rightarrow (\bra{00} + \bra{11})\otimes\bra{abc...} \\
    \bra{01}\otimes\bra{abc...} \rightarrow (\bra{01} + \bra{10})\otimes\bra{abc...} \\
    \bra{10}\otimes\bra{abc...} \rightarrow (\bra{00} - \bra{11})\otimes\bra{abc...} \\
    \bra{11}\otimes\bra{abc...} \rightarrow (\bra{01} - \bra{10})\otimes\bra{abc...}\\
\end{split}
\end{equation}

If we have CNOT H gates on every pair, we end up with all qubit pairs having entangled projections as above. Let us label:

\begin{equation}
\begin{split}
    \bra{S+} = \frac{1}{\sqrt{2}}(\bra{00} + \bra{11}) \\
    \bra{A+} = \frac{1}{\sqrt{2}}(\bra{01} + \bra{10}) \\
    \bra{S-} = \frac{1}{\sqrt{2}}(\bra{00} - \bra{11}) \\
    \bra{A-} = \frac{1}{\sqrt{2}}(\bra{01} - \bra{10}) \\
\end{split}
\end{equation}

The action of CNOT H creates several orbits, i.e. it will create entanglement between a few select states with each other. Specifically, for some fixed number of S and A states, the orbit will contain every combination of + and -. This is important because fixing the S and A nature of each qubit pair fixes the projected states whereas the +,- nature only affects relative sign.

We can write the linear combination of measurements in an example orbit of just S as:

\begin{equation}
\begin{split}
 \bra{\psi}(a\ket{S+} + b\ket{S-}\ket{\psi}
\end{split}
\end{equation}

We want this to equal $\braket{S_2}$. We can find coefficients by projecting onto the operator, for example:

\begin{equation}
    a = Tr(S_2 \ket{S+}\bra{S+}).
\end{equation}

Specifically, we see that:

\begin{equation}
\begin{split}
S_2\ket{S+} = \ket{S+}\\
S_2\ket{S-} = \ket{S-}\\
S_2\ket{A+} = \ket{A+}\\
S_2\ket{A-} = -\ket{A-}
\end{split}
\end{equation}

This means the coefficients are either 1 or -1 and, once they are solved, the linear combination of measurements can be made to estimate $\braket{S_2}$.

As an example, for $H_2$, the orbit of $\ket{0001}$ is $\ket{0001},\ket{0100},\ket{1011},\ket{1110}$. We can reorder the qubits and rewrite this as $\ket{0001},\ket{0010},\ket{1101},\ket{1110}$. This can be decomposed into:

\begin{equation}
\ket{S+}\ket{A+},\ket{S+}\ket{A-},\ket{S-}\ket{A+},\ket{S-}\ket{A-}
\end{equation}

Therefore, only the terms with $\ket{A-}$ will pick up a negative sign and we can write the expectation value is now:

\begin{equation}
\begin{split}
    \bra{\psi}(\ket{S+}\ket{A+}\bra{S+}\bra{A+} - \ket{S+}\ket{A-}\bra{S+}\bra{A-} + \\\ket{S-}\ket{A+}\bra{S-}\bra{A+} - \ket{S-}\ket{A-}\bra{S-}\bra{A-})\ket{\psi}
\end{split}
\end{equation}

We can also identify which measurements these correspond to by looking at the action of CNOT H on the states to be projected. Let $P_0$ be the probability of measuring outcome 0. We can now write the final expectation value as:

\begin{equation}
    P_{0001} - P_{0101} + P_{1001} - P_{1101}
\end{equation}

This process is then repeated for each orbit. This can be used for arbitrary numbers of qubits and serves as a low depth method of measuring the purity with two copies of a quantum circuit.

\subsection{Recreating Pauli Expectation Values}
While $\langle S_2 \rangle$ only requires pairwise CNOT gates and Hadamards on one copy of the ansatz, additional gates are needed to reconstruct the expectation value of arbitrary Pauli strings. Firstly, any X and Y observables can be mapped to Z by applying a Hadamard or Hadamard and $-\frac{\pi}{2}$ Phase gate respectively, so we only consider Pauli observables $O$ that contain $I$ or $Z$. Thus, $\langle S_2 O \rangle$, is only modified by adding a $-1$ phase to states where $O\ket{\psi} = -\ket{\psi}$. The additional gates we require must add this phase to those specific states. 

Unfortunately, in the general case, this will require multi-control-Z gates with control on arbitrary many qubits. Multi-control gates decompose into deep two qubit entangling gates and if even three qubit entangling gates are efficient, it may be better to simply measure expectation values with an ancilla and Hadamard test. However, the advantage depends on the topology and entangling to one ancilla may cause more swap gates than multi-control Z. 

\subsection{Decomposing Observables to Pauli Strings}
Direct decomposition of the observables into Pauli strings is possible and requires no additional entangling gates.

An observable, $O$, can be decomposed into Pauli strings by calculating the projection along each:
\begin{equation}
    a_i = \frac{1}{n} tr(P_i O).
\end{equation}

We can then write:
\begin{equation}
    O = \sum_{i} a_i P_i.
\end{equation}

As above, we can map measurements of Pauli strings with X and Y to Z by using a Hadamard or Hadamard and $-\frac{\pi}{2}$ phase gate on the corresponding qubit. Then the measurement sampling statistics can be combined with the relevant eigenvalues to get expectation values for every Pauli string with I or Z, since they commute. 

This procedure can be performed on the expectation value from equation~\ref{eq:fracti}. However, since these observables act on twice the qubit number, the corresponding Pauli decomposition will also act on twice the qubits. This means in the worst case scenario, it may take all $4^{2n}$ Pauli strings measurements to measure equation~\ref{eq:fracti}. 

As mentioned in the main text, a full tomography of one copy of the circuit is sufficient to get the density matrix and analytically perform virtual distillation up to arbitrary high copies. While this technique adds no extra entangling gates, the exponential scaling in measurements means it can only be used if the Pauli decomposition is guaranteed to be small.

\subsection{Ansatz Evaluation}

\begin{figure}
\includegraphics[clip,width=\columnwidth]{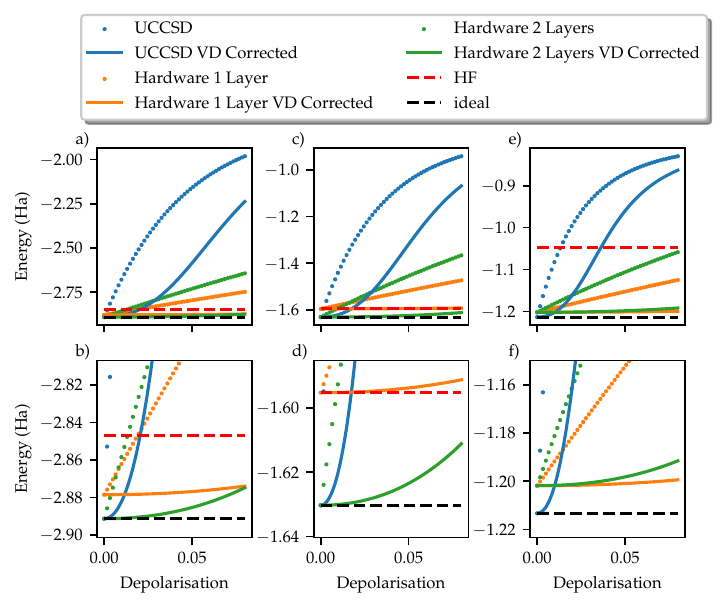}
\caption{Simulated Ansatz minimum energies raw and corrected with Virtual Distillation as depolarisation noise increases for UCCSD and Hardware Efficient Ansatz with one and two layers. a) shows the three qubit H$_3$ in Jordan-Wigner mapping and Z$_2$ symmetry reduction with expanded section in b). c) shows the four qubit H$_4$ with Parity mapping with expanded section in d). e) shows the four qubit H$_4$ with Jordan-Wigner mapping with expanded section in d). HF refers to Hartree-Fock initial guess and is the lowest depth circuit as it contains no entangling gates. Nuclear repulsion energy is not added.}
\label{fig:Ansatz}
\end{figure}

For three and four qubit calculations, we simulated UCCSD and Hardware Efficient Ansatz of increasing depth to determine the lowest depth Ansatz of appropriate accuracy to run on real devices. VQE was performed on noise free simulation with the basin-hopping algorithm to find the global minimum. Once the final parameters were found, the circuit was simulated with increasing depolarisation noise up to $0.1$ along with the full virtual distillation calculation and the results for all Ansatze are shown in Figure~\ref{fig:Ansatz}. 

UCCSD is able to find the ground state in all cases, however due to the circuit depth, depolarisation beyond $0.01$ causes it to be worse than the Hartree-Fock initial guess, even with VD corrections. Hardware Efficient Ansatz with one layer is not expressive enough to recover the ground state energy in any cases. It is also interesting to note that it is worse than the Hartree-Fock initial guess for four qubit Parity Hamiltonian shown in d). The Hardware Efficient Ansatz with two layers is able to recover the ground state and does not exceed the Hartree-Fock limit at 0.1 depolarisation for three qubit Hamiltonian in b) and the four qubit Parity Hamiltonian in d), however it is not expressive enough to find the ground state for the four qubit Jordan-Wigner Hamiltonian in f). As such, the obvious choice of Ansatz is the Hardware Efficient Ansatz with two layers for three qubit H$_3$ calculation and four qubit H$_2$ calculation with the Parity mapping. 

\subsection{Circuit Diagrams}

\begin{figure*}
    \centering
\scalebox{1.0}{
\Qcircuit @C=1.0em @R=0.2em @!R { \\
	 	\nghost{{q}_{0} :  } & \lstick{{q}_{0} :  } & \gate{\mathrm{R_Y}\,(\mathrm{\theta})} & \ctrl{1} & \qw & \qw\\
	 	\nghost{{q}_{1} :  } & \lstick{{q}_{1} :  } & \gate{\mathrm{X}} & \targ & \qw & \qw\\
\\ }}
    \caption{Manually reduced Ansatz for Z$_2$ symmetry reduced UCC H$_2$ two qubit Hamiltonian. The Hartree Fock initial guess is mapped to $\ket{01}$. The excitations were found to map to the space spanned by $\ket{01}$ and $\ket{10}$ with real coefficients.}
    \label{fig:app1}
\end{figure*}
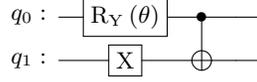

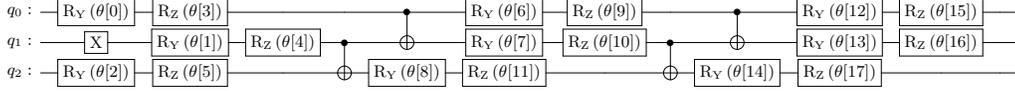
\begin{figure*}
    \centering\scalebox{0.7}{
\Qcircuit @C=1.0em @R=0.2em @!R { \\
	 	\nghost{{q}_{0} :  } & \lstick{{q}_{0} :  } & \gate{\mathrm{R_Y}\,(\mathrm{{\ensuremath{\theta}}[0]})} & \gate{\mathrm{R_Z}\,(\mathrm{{\ensuremath{\theta}}[3]})} & \qw & \qw & \ctrl{1} & \gate{\mathrm{R_Y}\,(\mathrm{{\ensuremath{\theta}}[6]})} & \gate{\mathrm{R_Z}\,(\mathrm{{\ensuremath{\theta}}[9]})} & \qw & \ctrl{1} & \gate{\mathrm{R_Y}\,(\mathrm{{\ensuremath{\theta}}[12]})} & \gate{\mathrm{R_Z}\,(\mathrm{{\ensuremath{\theta}}[15]})} & \qw & \qw\\
	 	\nghost{{q}_{1} :  } & \lstick{{q}_{1} :  } & \gate{\mathrm{X}} & \gate{\mathrm{R_Y}\,(\mathrm{{\ensuremath{\theta}}[1]})} & \gate{\mathrm{R_Z}\,(\mathrm{{\ensuremath{\theta}}[4]})} & \ctrl{1} & \targ & \gate{\mathrm{R_Y}\,(\mathrm{{\ensuremath{\theta}}[7]})} & \gate{\mathrm{R_Z}\,(\mathrm{{\ensuremath{\theta}}[10]})} & \ctrl{1} & \targ & \gate{\mathrm{R_Y}\,(\mathrm{{\ensuremath{\theta}}[13]})} & \gate{\mathrm{R_Z}\,(\mathrm{{\ensuremath{\theta}}[16]})} & \qw & \qw\\
	 	\nghost{{q}_{2} :  } & \lstick{{q}_{2} :  } & \gate{\mathrm{R_Y}\,(\mathrm{{\ensuremath{\theta}}[2]})} & \gate{\mathrm{R_Z}\,(\mathrm{{\ensuremath{\theta}}[5]})} & \qw & \targ & \gate{\mathrm{R_Y}\,(\mathrm{{\ensuremath{\theta}}[8]})} & \gate{\mathrm{R_Z}\,(\mathrm{{\ensuremath{\theta}}[11]})} & \qw & \targ & \gate{\mathrm{R_Y}\,(\mathrm{{\ensuremath{\theta}}[14]})} & \gate{\mathrm{R_Z}\,(\mathrm{{\ensuremath{\theta}}[17]})} & \qw & \qw & \qw\\
\\ }}
    \caption{Three qubit two layer Hardware efficient Ansatz for the Jordan-Wigner H$_3$ Hamiltonian. The initial Hartree-Fock guess is mapped to $\ket{010}$.}
    \label{fig:app2}
\end{figure*}

\begin{figure*}
\centering\scalebox{1.0}{\Qcircuit @C=1.0em @R=0.2em @!R { \\
	 	\nghost{{q}_{0} :  } & \lstick{{q}_{0} :  } & \gate{\mathrm{R_X}\,(\mathrm{\frac{-\pi}{2}})} & \gate{\mathrm{H}} & \ctrl{1} & \gate{\mathrm{R_X}\,(\mathrm{\frac{\pi}{4}})} & \ctrl{1} & \gate{\mathrm{H}} & \gate{\mathrm{R_X}\,(\mathrm{\frac{\pi}{2}})} & \qw & \qw\\
	 	\nghost{{q}_{1} :  } & \lstick{{q}_{1} :  } & \gate{\mathrm{R_X}\,(\mathrm{\frac{\pi}{2}})} & \qw & \targ & \gate{\mathrm{R_Z}\,(\mathrm{\frac{\pi}{4}})} & \targ & \gate{\mathrm{R_X}\,(\mathrm{\frac{-\pi}{2}})} & \qw & \qw & \qw\\
\\ }}

    \caption{B gate over two qubits. For larger number of qubits, we require B gates over every pair of qubits }
    \label{fig:app4}
\end{figure*}
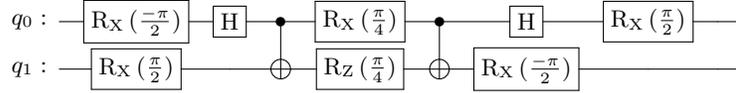

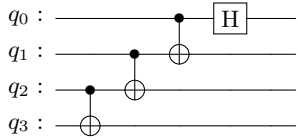
\begin{figure*}
    \centering\scalebox{1.0}{
\Qcircuit @C=1.0em @R=0.2em @!R { \\
	 	\nghost{{q}_{0} :  } & \lstick{{q}_{0} :  } & \qw & \qw & \ctrl{1} & \gate{\mathrm{H}} & \qw & \qw\\
	 	\nghost{{q}_{1} :  } & \lstick{{q}_{1} :  } & \qw & \ctrl{1} & \targ & \qw & \qw & \qw\\
	 	\nghost{{q}_{2} :  } & \lstick{{q}_{2} :  } & \ctrl{1} & \targ & \qw & \qw & \qw & \qw\\
	 	\nghost{{q}_{3} :  } & \lstick{{q}_{3} :  } & \targ & \qw & \qw & \qw & \qw & \qw\\
\\ }}
\caption{Entangling projection required to recreate Pauli sampling probabilities for H$_2$}
\label{fig:app5}
\end{figure*}

\begin{figure*}
    \centering\scalebox{1.0}{
\Qcircuit @C=1.0em @R=0.2em @!R { \\
	 	\nghost{{q}_{0} :  } & \lstick{{q}_{0} :  } & \ctrl{2} & \gate{\mathrm{H}} & \qw & \qw & \qw\\
	 	\nghost{{q}_{1} :  } & \lstick{{q}_{1} :  } & \qw & \ctrl{2} & \gate{\mathrm{H}} & \qw & \qw\\
	 	\nghost{{q}_{2} :  } & \lstick{{q}_{2} :  } & \targ & \qw & \qw & \qw & \qw\\
	 	\nghost{{q}_{3} :  } & \lstick{{q}_{3} :  } & \qw & \targ & \qw & \qw & \qw\\
\\ }}
\caption{Entangling projection required to recreate expectation value of $S_2$ for H$_2$}
\label{fig:app6}
\end{figure*}
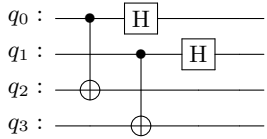

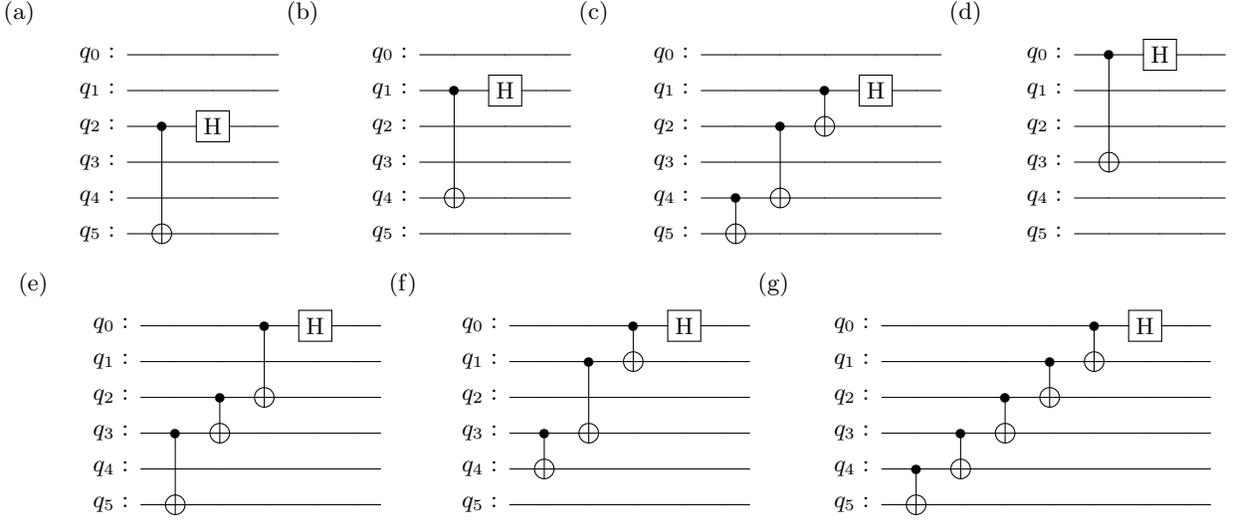
\begin{figure*}
    (a)\scalebox{1.0}{
\Qcircuit @C=1.0em @R=0.2em @!R { \\
	 	\nghost{{q}_{0} :  } & \lstick{{q}_{0} :  } & \qw & \qw & \qw & \qw\\
	 	\nghost{{q}_{1} :  } & \lstick{{q}_{1} :  } & \qw & \qw & \qw & \qw\\
	 	\nghost{{q}_{2} :  } & \lstick{{q}_{2} :  } & \ctrl{3} & \gate{\mathrm{H}} & \qw & \qw\\
	 	\nghost{{q}_{3} :  } & \lstick{{q}_{3} :  } & \qw & \qw & \qw & \qw\\
	 	\nghost{{q}_{4} :  } & \lstick{{q}_{4} :  } & \qw & \qw & \qw & \qw\\
	 	\nghost{{q}_{5} :  } & \lstick{{q}_{5} :  } & \targ & \qw & \qw & \qw\\
\\ }}
(b) \scalebox{1.0}{
\Qcircuit @C=1.0em @R=0.2em @!R { \\
	 	\nghost{{q}_{0} :  } & \lstick{{q}_{0} :  } & \qw & \qw & \qw & \qw\\
	 	\nghost{{q}_{1} :  } & \lstick{{q}_{1} :  } & \ctrl{3} & \gate{\mathrm{H}} & \qw & \qw\\
	 	\nghost{{q}_{2} :  } & \lstick{{q}_{2} :  } & \qw & \qw & \qw & \qw\\
	 	\nghost{{q}_{3} :  } & \lstick{{q}_{3} :  } & \qw & \qw & \qw & \qw\\
	 	\nghost{{q}_{4} :  } & \lstick{{q}_{4} :  } & \targ & \qw & \qw & \qw\\
	 	\nghost{{q}_{5} :  } & \lstick{{q}_{5} :  } & \qw & \qw & \qw & \qw\\
\\ }}
(c)\scalebox{1.0}{
\Qcircuit @C=1.0em @R=0.2em @!R { \\
	 	\nghost{{q}_{0} :  } & \lstick{{q}_{0} :  } & \qw & \qw & \qw & \qw & \qw & \qw\\
	 	\nghost{{q}_{1} :  } & \lstick{{q}_{1} :  } & \qw & \qw & \ctrl{1} & \gate{\mathrm{H}} & \qw & \qw\\
	 	\nghost{{q}_{2} :  } & \lstick{{q}_{2} :  } & \qw & \ctrl{2} & \targ & \qw & \qw & \qw\\
	 	\nghost{{q}_{3} :  } & \lstick{{q}_{3} :  } & \qw & \qw & \qw & \qw & \qw & \qw\\
	 	\nghost{{q}_{4} :  } & \lstick{{q}_{4} :  } & \ctrl{1} & \targ & \qw & \qw & \qw & \qw\\
	 	\nghost{{q}_{5} :  } & \lstick{{q}_{5} :  } & \targ & \qw & \qw & \qw & \qw & \qw\\
\\ }}
(d)\scalebox{1.0}{
\Qcircuit @C=1.0em @R=0.2em @!R { \\
	 	\nghost{{q}_{0} :  } & \lstick{{q}_{0} :  } & \ctrl{3} & \gate{\mathrm{H}} & \qw & \qw\\
	 	\nghost{{q}_{1} :  } & \lstick{{q}_{1} :  } & \qw & \qw & \qw & \qw\\
	 	\nghost{{q}_{2} :  } & \lstick{{q}_{2} :  } & \qw & \qw & \qw & \qw\\
	 	\nghost{{q}_{3} :  } & \lstick{{q}_{3} :  } & \targ & \qw & \qw & \qw\\
	 	\nghost{{q}_{4} :  } & \lstick{{q}_{4} :  } & \qw & \qw & \qw & \qw\\
	 	\nghost{{q}_{5} :  } & \lstick{{q}_{5} :  } & \qw & \qw & \qw & \qw\\
\\ }}
(e)\scalebox{1.0}{
\Qcircuit @C=1.0em @R=0.2em @!R { \\
	 	\nghost{{q}_{0} :  } & \lstick{{q}_{0} :  } & \qw & \qw & \ctrl{2} & \gate{\mathrm{H}} & \qw & \qw\\
	 	\nghost{{q}_{1} :  } & \lstick{{q}_{1} :  } & \qw & \qw & \qw & \qw & \qw & \qw\\
	 	\nghost{{q}_{2} :  } & \lstick{{q}_{2} :  } & \qw & \ctrl{1} & \targ & \qw & \qw & \qw\\
	 	\nghost{{q}_{3} :  } & \lstick{{q}_{3} :  } & \ctrl{2} & \targ & \qw & \qw & \qw & \qw\\
	 	\nghost{{q}_{4} :  } & \lstick{{q}_{4} :  } & \qw & \qw & \qw & \qw & \qw & \qw\\
	 	\nghost{{q}_{5} :  } & \lstick{{q}_{5} :  } & \targ & \qw & \qw & \qw & \qw & \qw\\
\\ }}
(f)\scalebox{1.0}{
\Qcircuit @C=1.0em @R=0.2em @!R { \\
	 	\nghost{{q}_{0} :  } & \lstick{{q}_{0} :  } & \qw & \qw & \ctrl{1} & \gate{\mathrm{H}} & \qw & \qw\\
	 	\nghost{{q}_{1} :  } & \lstick{{q}_{1} :  } & \qw & \ctrl{2} & \targ & \qw & \qw & \qw\\
	 	\nghost{{q}_{2} :  } & \lstick{{q}_{2} :  } & \qw & \qw & \qw & \qw & \qw & \qw\\
	 	\nghost{{q}_{3} :  } & \lstick{{q}_{3} :  } & \ctrl{1} & \targ & \qw & \qw & \qw & \qw\\
	 	\nghost{{q}_{4} :  } & \lstick{{q}_{4} :  } & \targ & \qw & \qw & \qw & \qw & \qw\\
	 	\nghost{{q}_{5} :  } & \lstick{{q}_{5} :  } & \qw & \qw & \qw & \qw & \qw & \qw\\
\\ }}
(g)\scalebox{1.0}{
\Qcircuit @C=1.0em @R=0.2em @!R { \\
	 	\nghost{{q}_{0} :  } & \lstick{{q}_{0} :  } & \qw & \qw & \qw & \qw & \ctrl{1} & \gate{\mathrm{H}} & \qw & \qw\\
	 	\nghost{{q}_{1} :  } & \lstick{{q}_{1} :  } & \qw & \qw & \qw & \ctrl{1} & \targ & \qw & \qw & \qw\\
	 	\nghost{{q}_{2} :  } & \lstick{{q}_{2} :  } & \qw & \qw & \ctrl{1} & \targ & \qw & \qw & \qw & \qw\\
	 	\nghost{{q}_{3} :  } & \lstick{{q}_{3} :  } & \qw & \ctrl{1} & \targ & \qw & \qw & \qw & \qw & \qw\\
	 	\nghost{{q}_{4} :  } & \lstick{{q}_{4} :  } & \ctrl{1} & \targ & \qw & \qw & \qw & \qw & \qw & \qw\\
	 	\nghost{{q}_{5} :  } & \lstick{{q}_{5} :  } & \targ & \qw & \qw & \qw & \qw & \qw & \qw & \qw\\
\\ }}
\caption{Entangling projections required to recreate Pauli sampling probabilities for H$_3$. Seven are required since the Pauli decomposition of the Hamiltonian includes all seven strings of the form $IIP, IPI, PII, IPP, PIP, PPI, PPP$, where $P \in {X,Y,Z}$. Pauli grouping or rotation can reduce the required number of circuits.}
\label{fig:app7}
\end{figure*}

\begin{figure*}
    (a)\scalebox{1.0}{
\Qcircuit @C=1.0em @R=0.2em @!R { \\
	 	\nghost{{q}_{0} :  } & \lstick{{q}_{0} :  } & \qw & \qw & \qw & \qw\\
	 	\nghost{{q}_{1} :  } & \lstick{{q}_{1} :  } & \qw & \qw & \qw & \qw\\
	 	\nghost{{q}_{2} :  } & \lstick{{q}_{2} :  } & \qw & \qw & \qw & \qw\\
	 	\nghost{{q}_{3} :  } & \lstick{{q}_{3} :  } & \ctrl{4} & \gate{\mathrm{H}} & \qw & \qw\\
	 	\nghost{{q}_{4} :  } & \lstick{{q}_{4} :  } & \qw & \qw & \qw & \qw\\
	 	\nghost{{q}_{5} :  } & \lstick{{q}_{5} :  } & \qw & \qw & \qw & \qw\\
	 	\nghost{{q}_{6} :  } & \lstick{{q}_{6} :  } & \qw & \qw & \qw & \qw\\
	 	\nghost{{q}_{7} :  } & \lstick{{q}_{7} :  } & \targ & \qw & \qw & \qw\\
\\ }}
(b)\scalebox{1.0}{
\Qcircuit @C=1.0em @R=0.2em @!R { \\
	 	\nghost{{q}_{0} :  } & \lstick{{q}_{0} :  } & \qw & \qw & \qw & \qw\\
	 	\nghost{{q}_{1} :  } & \lstick{{q}_{1} :  } & \qw & \qw & \qw & \qw\\
	 	\nghost{{q}_{2} :  } & \lstick{{q}_{2} :  } & \ctrl{4} & \gate{\mathrm{H}} & \qw & \qw\\
	 	\nghost{{q}_{3} :  } & \lstick{{q}_{3} :  } & \qw & \qw & \qw & \qw\\
	 	\nghost{{q}_{4} :  } & \lstick{{q}_{4} :  } & \qw & \qw & \qw & \qw\\
	 	\nghost{{q}_{5} :  } & \lstick{{q}_{5} :  } & \qw & \qw & \qw & \qw\\
	 	\nghost{{q}_{6} :  } & \lstick{{q}_{6} :  } & \targ & \qw & \qw & \qw\\
	 	\nghost{{q}_{7} :  } & \lstick{{q}_{7} :  } & \qw & \qw & \qw & \qw\\
\\ }}
(c)\scalebox{1.0}{
\Qcircuit @C=1.0em @R=0.2em @!R { \\
	 	\nghost{{q}_{0} :  } & \lstick{{q}_{0} :  } & \qw & \qw & \qw & \qw & \qw & \qw\\
	 	\nghost{{q}_{1} :  } & \lstick{{q}_{1} :  } & \qw & \qw & \qw & \qw & \qw & \qw\\
	 	\nghost{{q}_{2} :  } & \lstick{{q}_{2} :  } & \qw & \qw & \ctrl{1} & \gate{\mathrm{H}} & \qw & \qw\\
	 	\nghost{{q}_{3} :  } & \lstick{{q}_{3} :  } & \qw & \ctrl{3} & \targ & \qw & \qw & \qw\\
	 	\nghost{{q}_{4} :  } & \lstick{{q}_{4} :  } & \qw & \qw & \qw & \qw & \qw & \qw\\
	 	\nghost{{q}_{5} :  } & \lstick{{q}_{5} :  } & \qw & \qw & \qw & \qw & \qw & \qw\\
	 	\nghost{{q}_{6} :  } & \lstick{{q}_{6} :  } & \ctrl{1} & \targ & \qw & \qw & \qw & \qw\\
	 	\nghost{{q}_{7} :  } & \lstick{{q}_{7} :  } & \targ & \qw & \qw & \qw & \qw & \qw\\
\\ }}
(d)\scalebox{1.0}{
\Qcircuit @C=1.0em @R=0.2em @!R { \\
	 	\nghost{{q}_{0} :  } & \lstick{{q}_{0} :  } & \qw & \qw & \qw & \qw\\
	 	\nghost{{q}_{1} :  } & \lstick{{q}_{1} :  } & \ctrl{4} & \gate{\mathrm{H}} & \qw & \qw\\
	 	\nghost{{q}_{2} :  } & \lstick{{q}_{2} :  } & \qw & \qw & \qw & \qw\\
	 	\nghost{{q}_{3} :  } & \lstick{{q}_{3} :  } & \qw & \qw & \qw & \qw\\
	 	\nghost{{q}_{4} :  } & \lstick{{q}_{4} :  } & \qw & \qw & \qw & \qw\\
	 	\nghost{{q}_{5} :  } & \lstick{{q}_{5} :  } & \targ & \qw & \qw & \qw\\
	 	\nghost{{q}_{6} :  } & \lstick{{q}_{6} :  } & \qw & \qw & \qw & \qw\\
	 	\nghost{{q}_{7} :  } & \lstick{{q}_{7} :  } & \qw & \qw & \qw & \qw\\
\\ }}
(e)\scalebox{1.0}{
\Qcircuit @C=1.0em @R=0.2em @!R { \\
	 	\nghost{{q}_{0} :  } & \lstick{{q}_{0} :  } & \qw & \qw & \qw & \qw & \qw & \qw\\
	 	\nghost{{q}_{1} :  } & \lstick{{q}_{1} :  } & \qw & \qw & \ctrl{2} & \gate{\mathrm{H}} & \qw & \qw\\
	 	\nghost{{q}_{2} :  } & \lstick{{q}_{2} :  } & \qw & \qw & \qw & \qw & \qw & \qw\\
	 	\nghost{{q}_{3} :  } & \lstick{{q}_{3} :  } & \qw & \ctrl{2} & \targ & \qw & \qw & \qw\\
	 	\nghost{{q}_{4} :  } & \lstick{{q}_{4} :  } & \qw & \qw & \qw & \qw & \qw & \qw\\
	 	\nghost{{q}_{5} :  } & \lstick{{q}_{5} :  } & \ctrl{2} & \targ & \qw & \qw & \qw & \qw\\
	 	\nghost{{q}_{6} :  } & \lstick{{q}_{6} :  } & \qw & \qw & \qw & \qw & \qw & \qw\\
	 	\nghost{{q}_{7} :  } & \lstick{{q}_{7} :  } & \targ & \qw & \qw & \qw & \qw & \qw\\
\\ }}
(f)\scalebox{1.0}{
\Qcircuit @C=1.0em @R=0.2em @!R { \\
	 	\nghost{{q}_{0} :  } & \lstick{{q}_{0} :  } & \qw & \qw & \qw & \qw & \qw & \qw\\
	 	\nghost{{q}_{1} :  } & \lstick{{q}_{1} :  } & \qw & \qw & \ctrl{1} & \gate{\mathrm{H}} & \qw & \qw\\
	 	\nghost{{q}_{2} :  } & \lstick{{q}_{2} :  } & \qw & \ctrl{3} & \targ & \qw & \qw & \qw\\
	 	\nghost{{q}_{3} :  } & \lstick{{q}_{3} :  } & \qw & \qw & \qw & \qw & \qw & \qw\\
	 	\nghost{{q}_{4} :  } & \lstick{{q}_{4} :  } & \qw & \qw & \qw & \qw & \qw & \qw\\
	 	\nghost{{q}_{5} :  } & \lstick{{q}_{5} :  } & \ctrl{1} & \targ & \qw & \qw & \qw & \qw\\
	 	\nghost{{q}_{6} :  } & \lstick{{q}_{6} :  } & \targ & \qw & \qw & \qw & \qw & \qw\\
	 	\nghost{{q}_{7} :  } & \lstick{{q}_{7} :  } & \qw & \qw & \qw & \qw & \qw & \qw\\
\\ }}
(g)\scalebox{1.0}{
\Qcircuit @C=1.0em @R=0.2em @!R { \\
	 	\nghost{{q}_{0} :  } & \lstick{{q}_{0} :  } & \qw & \qw & \qw & \qw & \qw & \qw & \qw & \qw\\
	 	\nghost{{q}_{1} :  } & \lstick{{q}_{1} :  } & \qw & \qw & \qw & \qw & \ctrl{1} & \gate{\mathrm{H}} & \qw & \qw\\
	 	\nghost{{q}_{2} :  } & \lstick{{q}_{2} :  } & \qw & \qw & \qw & \ctrl{1} & \targ & \qw & \qw & \qw\\
	 	\nghost{{q}_{3} :  } & \lstick{{q}_{3} :  } & \qw & \qw & \ctrl{2} & \targ & \qw & \qw & \qw & \qw\\
	 	\nghost{{q}_{4} :  } & \lstick{{q}_{4} :  } & \qw & \qw & \qw & \qw & \qw & \qw & \qw & \qw\\
	 	\nghost{{q}_{5} :  } & \lstick{{q}_{5} :  } & \qw & \ctrl{1} & \targ & \qw & \qw & \qw & \qw & \qw\\
	 	\nghost{{q}_{6} :  } & \lstick{{q}_{6} :  } & \ctrl{1} & \targ & \qw & \qw & \qw & \qw & \qw & \qw\\
	 	\nghost{{q}_{7} :  } & \lstick{{q}_{7} :  } & \targ & \qw & \qw & \qw & \qw & \qw & \qw & \qw\\
\\ }}
(h)\scalebox{1.0}{
\Qcircuit @C=1.0em @R=0.2em @!R { \\
	 	\nghost{{q}_{0} :  } & \lstick{{q}_{0} :  } & \ctrl{4} & \gate{\mathrm{H}} & \qw & \qw\\
	 	\nghost{{q}_{1} :  } & \lstick{{q}_{1} :  } & \qw & \qw & \qw & \qw\\
	 	\nghost{{q}_{2} :  } & \lstick{{q}_{2} :  } & \qw & \qw & \qw & \qw\\
	 	\nghost{{q}_{3} :  } & \lstick{{q}_{3} :  } & \qw & \qw & \qw & \qw\\
	 	\nghost{{q}_{4} :  } & \lstick{{q}_{4} :  } & \targ & \qw & \qw & \qw\\
	 	\nghost{{q}_{5} :  } & \lstick{{q}_{5} :  } & \qw & \qw & \qw & \qw\\
	 	\nghost{{q}_{6} :  } & \lstick{{q}_{6} :  } & \qw & \qw & \qw & \qw\\
	 	\nghost{{q}_{7} :  } & \lstick{{q}_{7} :  } & \qw & \qw & \qw & \qw\\
\\ }}
(i)\scalebox{1.0}{
\Qcircuit @C=1.0em @R=0.2em @!R { \\
	 	\nghost{{q}_{0} :  } & \lstick{{q}_{0} :  } & \qw & \qw & \ctrl{3} & \gate{\mathrm{H}} & \qw & \qw\\
	 	\nghost{{q}_{1} :  } & \lstick{{q}_{1} :  } & \qw & \qw & \qw & \qw & \qw & \qw\\
	 	\nghost{{q}_{2} :  } & \lstick{{q}_{2} :  } & \qw & \qw & \qw & \qw & \qw & \qw\\
	 	\nghost{{q}_{3} :  } & \lstick{{q}_{3} :  } & \qw & \ctrl{1} & \targ & \qw & \qw & \qw\\
	 	\nghost{{q}_{4} :  } & \lstick{{q}_{4} :  } & \ctrl{3} & \targ & \qw & \qw & \qw & \qw\\
	 	\nghost{{q}_{5} :  } & \lstick{{q}_{5} :  } & \qw & \qw & \qw & \qw & \qw & \qw\\
	 	\nghost{{q}_{6} :  } & \lstick{{q}_{6} :  } & \qw & \qw & \qw & \qw & \qw & \qw\\
	 	\nghost{{q}_{7} :  } & \lstick{{q}_{7} :  } & \targ & \qw & \qw & \qw & \qw & \qw\\
\\ }}
(j)\scalebox{1.0}{
\Qcircuit @C=1.0em @R=0.2em @!R { \\
	 	\nghost{{q}_{0} :  } & \lstick{{q}_{0} :  } & \qw & \qw & \ctrl{2} & \gate{\mathrm{H}} & \qw & \qw\\
	 	\nghost{{q}_{1} :  } & \lstick{{q}_{1} :  } & \qw & \qw & \qw & \qw & \qw & \qw\\
	 	\nghost{{q}_{2} :  } & \lstick{{q}_{2} :  } & \qw & \ctrl{2} & \targ & \qw & \qw & \qw\\
	 	\nghost{{q}_{3} :  } & \lstick{{q}_{3} :  } & \qw & \qw & \qw & \qw & \qw & \qw\\
	 	\nghost{{q}_{4} :  } & \lstick{{q}_{4} :  } & \ctrl{2} & \targ & \qw & \qw & \qw & \qw\\
	 	\nghost{{q}_{5} :  } & \lstick{{q}_{5} :  } & \qw & \qw & \qw & \qw & \qw & \qw\\
	 	\nghost{{q}_{6} :  } & \lstick{{q}_{6} :  } & \targ & \qw & \qw & \qw & \qw & \qw\\
	 	\nghost{{q}_{7} :  } & \lstick{{q}_{7} :  } & \qw & \qw & \qw & \qw & \qw & \qw\\
\\ }}
(k)\scalebox{1.0}{
\Qcircuit @C=1.0em @R=0.2em @!R { \\
	 	\nghost{{q}_{0} :  } & \lstick{{q}_{0} :  } & \qw & \qw & \qw & \qw & \ctrl{2} & \gate{\mathrm{H}} & \qw & \qw\\
	 	\nghost{{q}_{1} :  } & \lstick{{q}_{1} :  } & \qw & \qw & \qw & \qw & \qw & \qw & \qw & \qw\\
	 	\nghost{{q}_{2} :  } & \lstick{{q}_{2} :  } & \qw & \qw & \qw & \ctrl{1} & \targ & \qw & \qw & \qw\\
	 	\nghost{{q}_{3} :  } & \lstick{{q}_{3} :  } & \qw & \qw & \ctrl{1} & \targ & \qw & \qw & \qw & \qw\\
	 	\nghost{{q}_{4} :  } & \lstick{{q}_{4} :  } & \qw & \ctrl{2} & \targ & \qw & \qw & \qw & \qw & \qw\\
	 	\nghost{{q}_{5} :  } & \lstick{{q}_{5} :  } & \qw & \qw & \qw & \qw & \qw & \qw & \qw & \qw\\
	 	\nghost{{q}_{6} :  } & \lstick{{q}_{6} :  } & \ctrl{1} & \targ & \qw & \qw & \qw & \qw & \qw & \qw\\
	 	\nghost{{q}_{7} :  } & \lstick{{q}_{7} :  } & \targ & \qw & \qw & \qw & \qw & \qw & \qw & \qw\\
\\ }}
(l)\scalebox{1.0}{
\Qcircuit @C=1.0em @R=0.2em @!R { \\
	 	\nghost{{q}_{0} :  } & \lstick{{q}_{0} :  } & \qw & \qw & \ctrl{1} & \gate{\mathrm{H}} & \qw & \qw\\
	 	\nghost{{q}_{1} :  } & \lstick{{q}_{1} :  } & \qw & \ctrl{3} & \targ & \qw & \qw & \qw\\
	 	\nghost{{q}_{2} :  } & \lstick{{q}_{2} :  } & \qw & \qw & \qw & \qw & \qw & \qw\\
	 	\nghost{{q}_{3} :  } & \lstick{{q}_{3} :  } & \qw & \qw & \qw & \qw & \qw & \qw\\
	 	\nghost{{q}_{4} :  } & \lstick{{q}_{4} :  } & \ctrl{1} & \targ & \qw & \qw & \qw & \qw\\
	 	\nghost{{q}_{5} :  } & \lstick{{q}_{5} :  } & \targ & \qw & \qw & \qw & \qw & \qw\\
	 	\nghost{{q}_{6} :  } & \lstick{{q}_{6} :  } & \qw & \qw & \qw & \qw & \qw & \qw\\
	 	\nghost{{q}_{7} :  } & \lstick{{q}_{7} :  } & \qw & \qw & \qw & \qw & \qw & \qw\\
\\ }}
(m)\scalebox{1.0}{
\Qcircuit @C=1.0em @R=0.2em @!R { \\
	 	\nghost{{q}_{0} :  } & \lstick{{q}_{0} :  } & \qw & \qw & \qw & \qw & \ctrl{1} & \gate{\mathrm{H}} & \qw & \qw\\
	 	\nghost{{q}_{1} :  } & \lstick{{q}_{1} :  } & \qw & \qw & \qw & \ctrl{2} & \targ & \qw & \qw & \qw\\
	 	\nghost{{q}_{2} :  } & \lstick{{q}_{2} :  } & \qw & \qw & \qw & \qw & \qw & \qw & \qw & \qw\\
	 	\nghost{{q}_{3} :  } & \lstick{{q}_{3} :  } & \qw & \qw & \ctrl{1} & \targ & \qw & \qw & \qw & \qw\\
	 	\nghost{{q}_{4} :  } & \lstick{{q}_{4} :  } & \qw & \ctrl{1} & \targ & \qw & \qw & \qw & \qw & \qw\\
	 	\nghost{{q}_{5} :  } & \lstick{{q}_{5} :  } & \ctrl{2} & \targ & \qw & \qw & \qw & \qw & \qw & \qw\\
	 	\nghost{{q}_{6} :  } & \lstick{{q}_{6} :  } & \qw & \qw & \qw & \qw & \qw & \qw & \qw & \qw\\
	 	\nghost{{q}_{7} :  } & \lstick{{q}_{7} :  } & \targ & \qw & \qw & \qw & \qw & \qw & \qw & \qw\\
\\ }}
(n)\scalebox{1.0}{
\Qcircuit @C=1.0em @R=0.2em @!R { \\
	 	\nghost{{q}_{0} :  } & \lstick{{q}_{0} :  } & \qw & \qw & \qw & \qw & \ctrl{1} & \gate{\mathrm{H}} & \qw & \qw\\
	 	\nghost{{q}_{1} :  } & \lstick{{q}_{1} :  } & \qw & \qw & \qw & \ctrl{1} & \targ & \qw & \qw & \qw\\
	 	\nghost{{q}_{2} :  } & \lstick{{q}_{2} :  } & \qw & \qw & \ctrl{2} & \targ & \qw & \qw & \qw & \qw\\
	 	\nghost{{q}_{3} :  } & \lstick{{q}_{3} :  } & \qw & \qw & \qw & \qw & \qw & \qw & \qw & \qw\\
	 	\nghost{{q}_{4} :  } & \lstick{{q}_{4} :  } & \qw & \ctrl{1} & \targ & \qw & \qw & \qw & \qw & \qw\\
	 	\nghost{{q}_{5} :  } & \lstick{{q}_{5} :  } & \ctrl{1} & \targ & \qw & \qw & \qw & \qw & \qw & \qw\\
	 	\nghost{{q}_{6} :  } & \lstick{{q}_{6} :  } & \targ & \qw & \qw & \qw & \qw & \qw & \qw & \qw\\
	 	\nghost{{q}_{7} :  } & \lstick{{q}_{7} :  } & \qw & \qw & \qw & \qw & \qw & \qw & \qw & \qw\\
\\ }}
(o)\scalebox{1.0}{
\Qcircuit @C=1.0em @R=0.2em @!R { \\
	 	\nghost{{q}_{0} :  } & \lstick{{q}_{0} :  } & \qw & \qw & \qw & \qw & \qw & \qw & \ctrl{1} & \gate{\mathrm{H}} & \qw & \qw\\
	 	\nghost{{q}_{1} :  } & \lstick{{q}_{1} :  } & \qw & \qw & \qw & \qw & \qw & \ctrl{1} & \targ & \qw & \qw & \qw\\
	 	\nghost{{q}_{2} :  } & \lstick{{q}_{2} :  } & \qw & \qw & \qw & \qw & \ctrl{1} & \targ & \qw & \qw & \qw & \qw\\
	 	\nghost{{q}_{3} :  } & \lstick{{q}_{3} :  } & \qw & \qw & \qw & \ctrl{1} & \targ & \qw & \qw & \qw & \qw & \qw\\
	 	\nghost{{q}_{4} :  } & \lstick{{q}_{4} :  } & \qw & \qw & \ctrl{1} & \targ & \qw & \qw & \qw & \qw & \qw & \qw\\
	 	\nghost{{q}_{5} :  } & \lstick{{q}_{5} :  } & \qw & \ctrl{1} & \targ & \qw & \qw & \qw & \qw & \qw & \qw & \qw\\
	 	\nghost{{q}_{6} :  } & \lstick{{q}_{6} :  } & \ctrl{1} & \targ & \qw & \qw & \qw & \qw & \qw & \qw & \qw & \qw\\
	 	\nghost{{q}_{7} :  } & \lstick{{q}_{7} :  } & \targ & \qw & \qw & \qw & \qw & \qw & \qw & \qw & \qw & \qw\\
\\ }}
\caption{Entangling projections required to recreate Pauli sampling probabilities for H$_2$ mapped to four qubits. All fifteen are required due to the Pauli decomposition of the Hamiltonian. Pauli grouping or rotation can reduce the required number of circuits.}
\label{fig:app9}
\end{figure*}
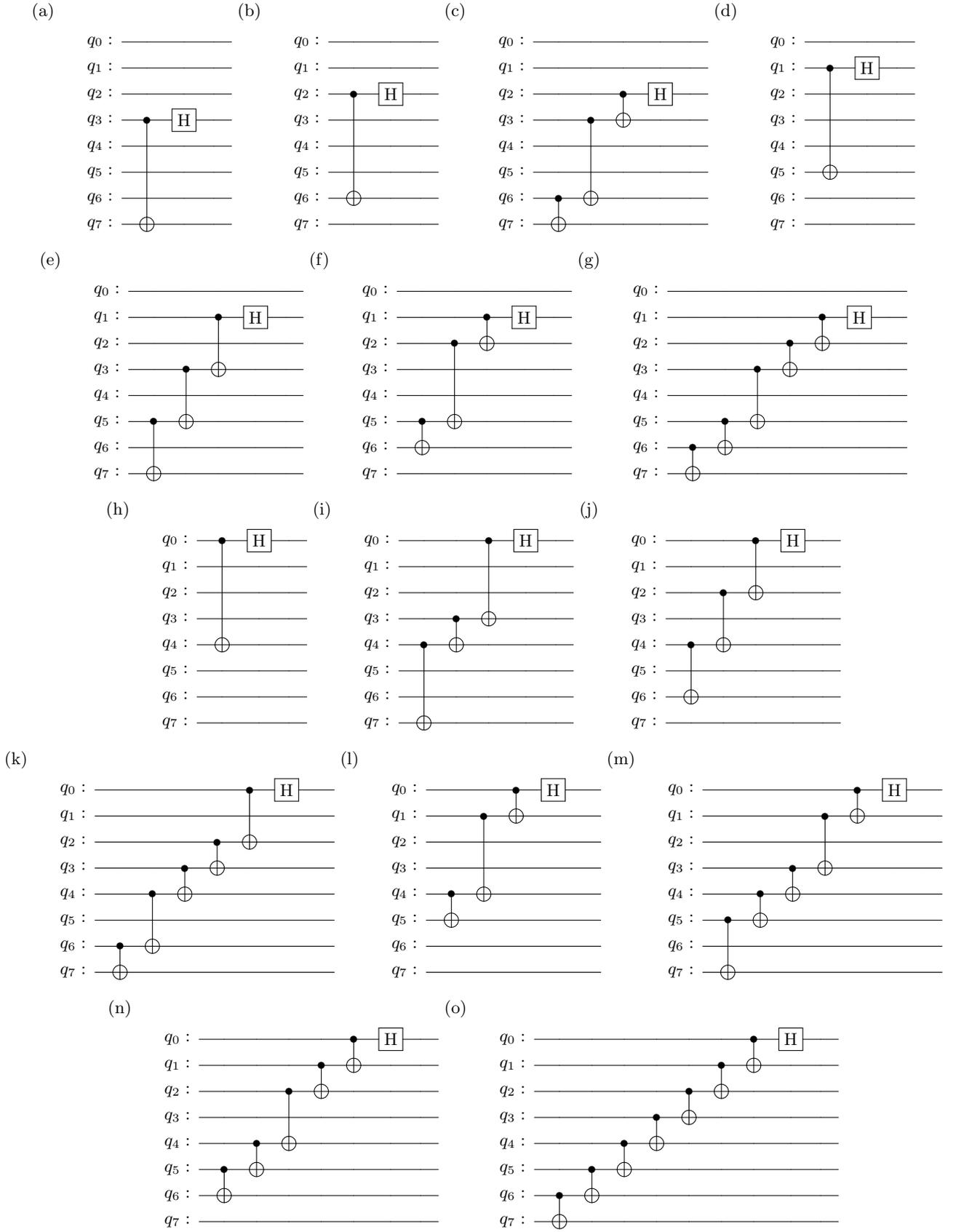

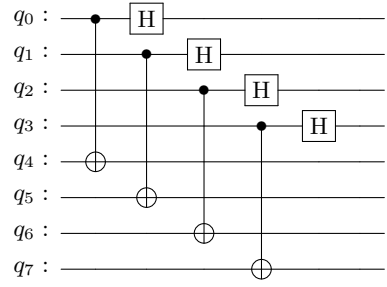
\begin{figure*}
    \centering\scalebox{1.0}{
\Qcircuit @C=1.0em @R=0.2em @!R { \\
	 	\nghost{{q}_{0} :  } & \lstick{{q}_{0} :  } & \ctrl{4} & \gate{\mathrm{H}} & \qw & \qw & \qw & \qw & \qw\\
	 	\nghost{{q}_{1} :  } & \lstick{{q}_{1} :  } & \qw & \ctrl{4} & \gate{\mathrm{H}} & \qw & \qw & \qw & \qw\\
	 	\nghost{{q}_{2} :  } & \lstick{{q}_{2} :  } & \qw & \qw & \ctrl{4} & \gate{\mathrm{H}} & \qw & \qw & \qw\\
	 	\nghost{{q}_{3} :  } & \lstick{{q}_{3} :  } & \qw & \qw & \qw & \ctrl{4} & \gate{\mathrm{H}} & \qw & \qw\\
	 	\nghost{{q}_{4} :  } & \lstick{{q}_{4} :  } & \targ & \qw & \qw & \qw & \qw & \qw & \qw\\
	 	\nghost{{q}_{5} :  } & \lstick{{q}_{5} :  } & \qw & \targ & \qw & \qw & \qw & \qw & \qw\\
	 	\nghost{{q}_{6} :  } & \lstick{{q}_{6} :  } & \qw & \qw & \targ & \qw & \qw & \qw & \qw\\
	 	\nghost{{q}_{7} :  } & \lstick{{q}_{7} :  } & \qw & \qw & \qw & \targ & \qw & \qw & \qw\\
\\ }}
\caption{Entangling projection required to recreate expectation value of $S_2$ for H$_2$ mapped to four qubits}
\label{fig:app10}
\end{figure*}

In this section we will show the circuits run on real devices in this work. First we show the ansatz used. For H$_2$ (two qubit) shown in figure~\ref{fig:app1} we design our own ansatz based on the symmetry of the ground state in the parity mapping with Z$_2$ symmetry reduction. For H$_3$ (three qubit) and H$_2$ (four qubit) we use a basic SU2 hardware efficient circuit with two repetitions in figures~\ref{fig:app2} and \ref{fig:app3}. Next we show the B gates to directly measure the $S_2$ expectation value in Figure\ref{fig:app4}. This is applied to every pair of qubits. For this work, we replace the B gates with combinations of computational and entangled measurements. The entangled gates for Pauli sampling reconstructing of H$_2$ (two qubit) are given in figure~\ref{fig:app5}. The entangling circuit for reconstructing the $S_2$ expectation value is given in figure~\ref{fig:app6}. Similarly, the entangled gates for H$_3$ (three qubits) Paulis are given in figure~\ref{fig:app7} and $S_2$ in figure~\ref{fig:app8}. The entangled gates for H$_2$ (four qubits) are given in figure~\ref{fig:app9} and $S_2$ in figure~\ref{fig:app10}.

\begin{figure*}
    \scalebox{0.7}{
\Qcircuit @C=1.0em @R=0.2em @!R { \\
	 	\nghost{{q}_{0} :  } & \lstick{{q}_{0} :  } & \gate{\mathrm{X}} & \gate{\mathrm{R_Y}\,(\mathrm{{\ensuremath{\theta}}[0]})} & \gate{\mathrm{R_Z}\,(\mathrm{{\ensuremath{\theta}}[4]})} & \qw & \ctrl{1} & \gate{\mathrm{R_Y}\,(\mathrm{{\ensuremath{\theta}}[8]})} & \gate{\mathrm{R_Z}\,(\mathrm{{\ensuremath{\theta}}[12]})} & \qw & \ctrl{1} & \gate{\mathrm{R_Y}\,(\mathrm{{\ensuremath{\theta}}[16]})} & \gate{\mathrm{R_Z}\,(\mathrm{{\ensuremath{\theta}}[20]})} & \qw & \qw\\
	 	\nghost{{q}_{1} :  } & \lstick{{q}_{1} :  } & \gate{\mathrm{X}} & \gate{\mathrm{R_Y}\,(\mathrm{{\ensuremath{\theta}}[1]})} & \gate{\mathrm{R_Z}\,(\mathrm{{\ensuremath{\theta}}[5]})} & \ctrl{1} & \targ & \gate{\mathrm{R_Y}\,(\mathrm{{\ensuremath{\theta}}[9]})} & \gate{\mathrm{R_Z}\,(\mathrm{{\ensuremath{\theta}}[13]})} & \ctrl{1} & \targ & \gate{\mathrm{R_Y}\,(\mathrm{{\ensuremath{\theta}}[17]})} & \gate{\mathrm{R_Z}\,(\mathrm{{\ensuremath{\theta}}[21]})} & \qw & \qw\\
	 	\nghost{{q}_{2} :  } & \lstick{{q}_{2} :  } & \gate{\mathrm{R_Y}\,(\mathrm{{\ensuremath{\theta}}[2]})} & \gate{\mathrm{R_Z}\,(\mathrm{{\ensuremath{\theta}}[6]})} & \ctrl{1} & \targ & \gate{\mathrm{R_Y}\,(\mathrm{{\ensuremath{\theta}}[10]})} & \gate{\mathrm{R_Z}\,(\mathrm{{\ensuremath{\theta}}[14]})} & \ctrl{1} & \targ & \gate{\mathrm{R_Y}\,(\mathrm{{\ensuremath{\theta}}[18]})} & \gate{\mathrm{R_Z}\,(\mathrm{{\ensuremath{\theta}}[22]})} & \qw & \qw & \qw\\
	 	\nghost{{q}_{3} :  } & \lstick{{q}_{3} :  } & \gate{\mathrm{R_Y}\,(\mathrm{{\ensuremath{\theta}}[3]})} & \gate{\mathrm{R_Z}\,(\mathrm{{\ensuremath{\theta}}[7]})} & \targ & \gate{\mathrm{R_Y}\,(\mathrm{{\ensuremath{\theta}}[11]})} & \gate{\mathrm{R_Z}\,(\mathrm{{\ensuremath{\theta}}[15]})} & \qw & \targ & \gate{\mathrm{R_Y}\,(\mathrm{{\ensuremath{\theta}}[19]})} & \gate{\mathrm{R_Z}\,(\mathrm{{\ensuremath{\theta}}[23]})} & \qw & \qw & \qw & \qw\\
\\ }}
\caption{Four qubit two layer hardware efficient Ansatz for H$_2$ molecule with Parity mapping but no symmetry reduction. Hartree Fock state is mapped to $\ket{1100}$.}
\label{fig:app3}
\end{figure*}
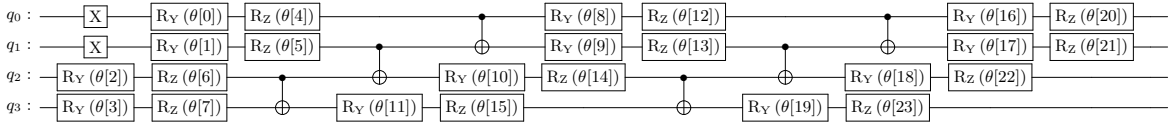

\begin{figure*}
    \centering\scalebox{1.0}{
\Qcircuit @C=1.0em @R=0.2em @!R { \\
	 	\nghost{{q}_{0} :  } & \lstick{{q}_{0} :  } & \ctrl{3} & \gate{\mathrm{H}} & \qw & \qw & \qw & \qw\\
	 	\nghost{{q}_{1} :  } & \lstick{{q}_{1} :  } & \qw & \ctrl{3} & \gate{\mathrm{H}} & \qw & \qw & \qw\\
	 	\nghost{{q}_{2} :  } & \lstick{{q}_{2} :  } & \qw & \qw & \ctrl{3} & \gate{\mathrm{H}} & \qw & \qw\\
	 	\nghost{{q}_{3} :  } & \lstick{{q}_{3} :  } & \targ & \qw & \qw & \qw & \qw & \qw\\
	 	\nghost{{q}_{4} :  } & \lstick{{q}_{4} :  } & \qw & \targ & \qw & \qw & \qw & \qw\\
	 	\nghost{{q}_{5} :  } & \lstick{{q}_{5} :  } & \qw & \qw & \targ & \qw & \qw & \qw\\
\\ }}
\caption{Entangling projection required to recreate expectation value of $S_2$ for H$_3$}
\label{fig:app8}
\end{figure*}
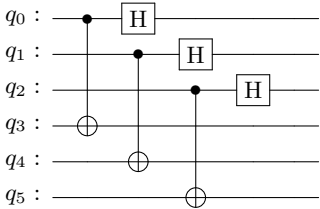

\begin{figure*}
    \centering
    \includegraphics[width=\columnwidth]{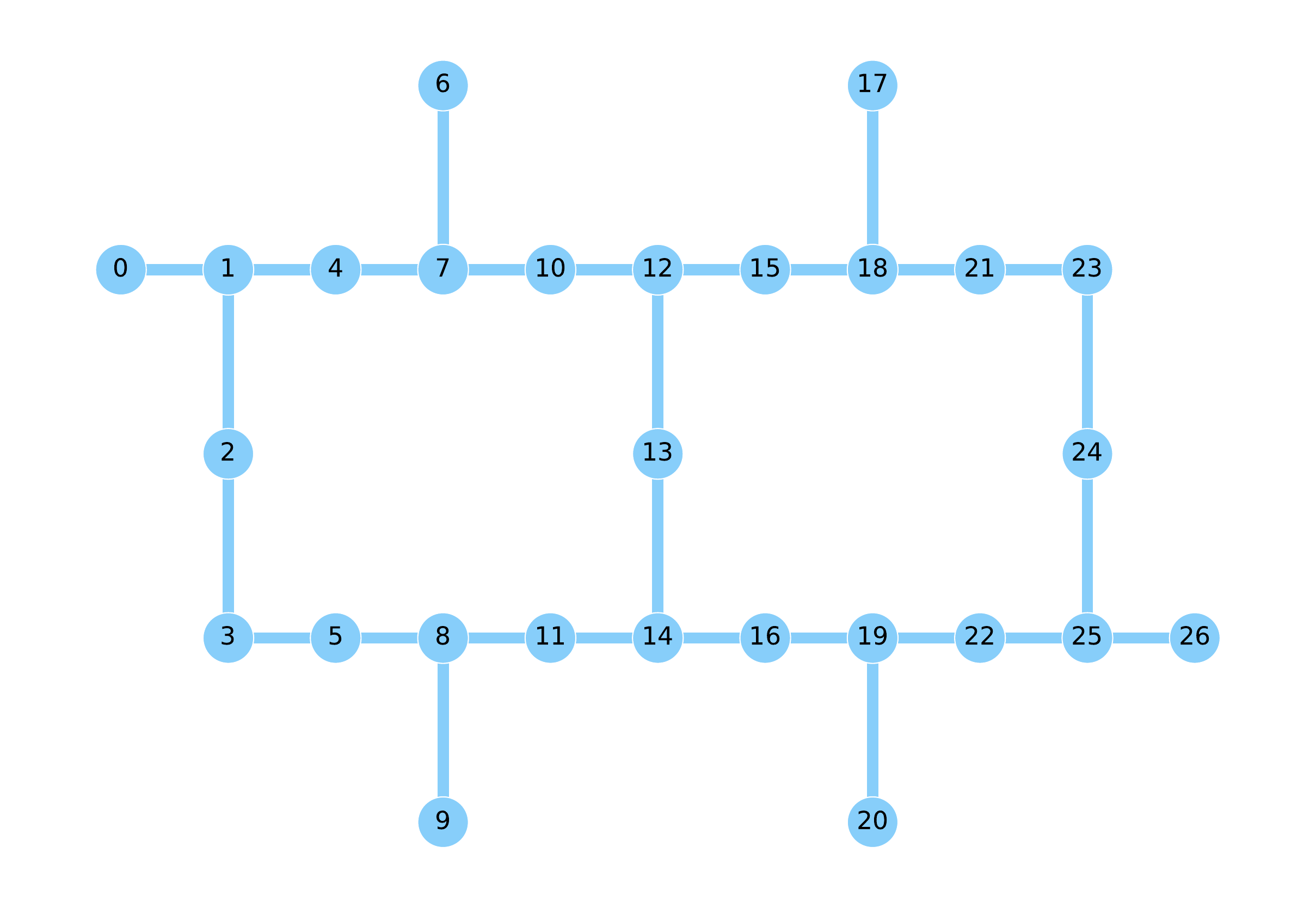}
    \caption{Layout of the $ibm\_hanoi$ device. Two qubit calculation uses qubits 1,4,7,8. Three-qubit calculation uses 5,3,2,1,4,7. Four-qubit calculation uses 1,2,3,6,4,7,5,8.}
    \label{fig:enter-label}
\end{figure*}

\subsection{IBM Device Characteristics}

In this section, we show the layout of the $ibm\_hanoi$ chip used for demonstrations in this work in figure~\ref{fig:enter-label} along with noise spectroscopy collected around the time of the simulations in table~\ref{table}. It should be noted that other sources of noise exist in real devices and this only represents the sources of noise that were measured. 

\afterpage{
\clearpage
\thispagestyle{empty}
\centering
\begin{sidewaystable}
\caption{Table of $ibm\_hanoi$ device parameters. The frequency, T$_1$, T$_2$, and readout noise are given for each qubit and the CX error is given for pairs of qubits.}

\adjustbox{scale=0.88,right=25cm,vspace*=0pt -10cm}{
\begin{tabular}
{|l|l|l|l|l|l|l|l|l|l|l|l|l|l|l|l|l|l|l|l|l|l|l|l|l|l|l|l|}

\hline 
& $q_{0}$ & $q_{1}$ & $q_{2}$ & $q_{3}$ & $q_{4}$ & $q_{5}$ & $q_{6}$ & $q_{7}$ & $q_{8}$ & $q_{9}$ & $q_{10}$ & $q_{11}$ & $q_{12}$ & $q_{13}$ & $q_{14}$ & $q_{15}$ & $q_{16}$ & $q_{17}$ & $q_{18}$ & $q_{19}$ & $q_{20}$ & $q_{21}$ & $q_{22}$ & $q_{23}$ & $q_{24}$ & $q_{25}$ & $q_{26}$\\
\hline 
Frequency (GHz) & $5.035$ & $5.155$ & $5.256$ & $5.097$ & $5.072$ & $5.206$ & $5.020$ & $4.919$ & $5.031$ & $4.874$ & $4.821$ & $5.162$ & $4.718$ & $4.963$ & $5.046$ & $4.923$ & $4.883$ & $5.222$ & $4.967$ & $5.002$ & $5.095$ & $4.839$ & $4.919$ & $4.920$ & $4.991$ & $4.812$ & $5.019$\\ \hline 
T$_1$($\mu$ s) & $174.9$ & $224.6$ & $127.8$ & $103.4$ & $134.0$ & $179.8$ & $132.2$ & $170.0$ & $103.7$ & $135.1$ & $114.3$ & $119.3$ & $183.7$ & $249.7$ & $104.3$ & $164.7$ & $167.8$ & $119.4$ & $123.0$ & $145.7$ & $66.70$ & $158.4$ & $80.58$ & $166.4$ & $136.9$ & $175.7$ & $83.87$\\ \hline 
T$_2$($\mu$ s) & $210.5$ & $159.4$ & $183.0$ & $32.45$ & $14.13$ & $231.8$ & $278.4$ & $125.8$ & $185.2$ & $114.7$ & $102.6$ & $243.0$ & $160.8$ & $114.3$ & $23.30$ & $37.32$ & $228.6$ & $67.57$ & $162.9$ & $126.8$ & $93.77$ & $36.04$ & $88.66$ & $241.7$ & $27.89$ & $108.9$ & $31.50$\\ \hline 
Readout Error & $0.0082$ & $0.0098$ & $0.0121$ & $0.0083$ & $0.0078$ & $0.0087$ & $0.046$ & $0.0105$ & $0.0106$ & $0.0072$ & $0.0121$ & $0.088$ & $0.146$ & $0.068$ & $0.0103$ & $0.0521$ & $0.0119$ & $0.0178$ & $0.0154$ & $0.0073$ & $0.0100$ & $0.0271$ & $0.0125$ & $0.0127$ & $0.0116$ & $0.0080$ & $0.0080$\\ \hline 
CX error $q_{0}$&  & $0.52$ &   &   &   &   &   &   &   &   &   &   &   &   &   &   &   &   &   &   &   &   &   &   &   &   &  \\ \hline CX error $q_{1}$&$0.52$ &   & $0.30$ &   & $0.38$ &   &   &   &   &   &   &   &   &   &   &   &   &   &   &   &   &   &   &   &   &   &  \\ \hline CX error $q_{2}$&  & $0.30$ &   & $0.95$ &   &   &   &   &   &   &   &   &   &   &   &   &   &   &   &   &   &   &   &   &   &   &  \\ \hline CX error $q_{3}$&  &   & $0.95$ &   &   & $0.37$ &   &   &   &   &   &   &   &   &   &   &   &   &   &   &   &   &   &   &   &   &  \\ \hline CX error $q_{4}$&  & $0.38$ &   &   &   &   &   & $1.40$ &   &   &   &   &   &   &   &   &   &   &   &   &   &   &   &   &   &   &  \\ \hline CX error $q_{5}$&  &   &   & $0.37$ &   &   &   &   & $100$ &   &   &   &   &   &   &   &   &   &   &   &   &   &   &   &   &   &  \\ \hline CX error $q_{6}$&  &   &   &   &   &   &   & $0.71$ &   &   &   &   &   &   &   &   &   &   &   &   &   &   &   &   &   &   &  \\ \hline CX error $q_{7}$&  &   &   &   & $1.40$ &   & $0.71$ &   &   &   & $0.67$ &   &   &   &   &   &   &   &   &   &   &   &   &   &   &   &  \\ \hline CX error $q_{8}$&  &   &   &   &   & $100$ &   &   &   & $1.03$ &   & $0.55$ &   &   &   &   &   &   &   &   &   &   &   &   &   &   &  \\ \hline CX error $q_{9}$&  &   &   &   &   &   &   &   & $1.03$ &   &   &   &   &   &   &   &   &   &   &   &   &   &   &   &   &   &  \\ \hline CX error $q_{10}$&  &   &   &   &   &   &   & $0.67$ &   &   &   &   & $0.60$ &   &   &   &   &   &   &   &   &   &   &   &   &   &  \\ \hline CX error $q_{11}$&  &   &   &   &   &   &   &   & $0.55$ &   &   &   &   &   & $0.72$ &   &   &   &   &   &   &   &   &   &   &   &  \\ \hline CX error $q_{12}$&  &   &   &   &   &   &   &   &   &   & $0.60$ &   &   & $0.53$ &   & $0.88$ &   &   &   &   &   &   &   &   &   &   &  \\ \hline CX error $q_{13}$&  &   &   &   &   &   &   &   &   &   &   &   & $0.53$ &   & $0.55$ &   &   &   &   &   &   &   &   &   &   &   &  \\ \hline CX error $q_{14}$&  &   &   &   &   &   &   &   &   &   &   & $0.72$ &   & $0.55$ &   &   & $1.32$ &   &   &   &   &   &   &   &   &   &  \\ \hline CX error $q_{15}$&  &   &   &   &   &   &   &   &   &   &   &   & $0.88$ &   &   &   &   &   & $1.22$ &   &   &   &   &   &   &   &  \\ \hline CX error $q_{16}$&  &   &   &   &   &   &   &   &   &   &   &   &   &   & $1.32$ &   &   &   &   & $0.38$ &   &   &   &   &   &   &  \\ \hline CX error $q_{17}$&  &   &   &   &   &   &   &   &   &   &   &   &   &   &   &   &   &   & $0.57$ &   &   &   &   &   &   &   &  \\ \hline CX error $q_{18}$&  &   &   &   &   &   &   &   &   &   &   &   &   &   &   & $1.22$ &   & $0.57$ &   &   &   & $0.40$ &   &   &   &   &  \\ \hline CX error $q_{19}$&  &   &   &   &   &   &   &   &   &   &   &   &   &   &   &   & $0.38$ &   &   &   & $100$ &   & $0.69$ &   &   &   &  \\ \hline CX error $q_{20}$&  &   &   &   &   &   &   &   &   &   &   &   &   &   &   &   &   &   &   & $100$ &   &   &   &   &   &   &  \\ \hline CX error $q_{21}$&  &   &   &   &   &   &   &   &   &   &   &   &   &   &   &   &   &   & $0.40$ &   &   &   &   & $0.66$ &   &   &  \\ \hline CX error $q_{22}$&  &   &   &   &   &   &   &   &   &   &   &   &   &   &   &   &   &   &   & $0.69$ &   &   &   &   &   & $0.83$ &  \\ \hline CX error $q_{23}$&  &   &   &   &   &   &   &   &   &   &   &   &   &   &   &   &   &   &   &   &   & $0.66$ &   &   & $1.54$ &   &  \\ \hline CX error $q_{24}$&  &   &   &   &   &   &   &   &   &   &   &   &   &   &   &   &   &   &   &   &   &   &   & $1.54$ &   & $2.54$ &  \\ \hline CX error $q_{25}$&  &   &   &   &   &   &   &   &   &   &   &   &   &   &   &   &   &   &   &   &   &   & $0.83$ &   & $2.54$ &   & $0.68$\\ \hline CX error $q_{26}$&  &   &   &   &   &   &   &   &   &   &   &   &   &   &   &   &   &   &   &   &   &   &   &   &   & $0.68$ &  \\ \hline  
\end{tabular}
}\label{table}
\end{sidewaystable}
}

\end{document}